\documentclass[11pt,a4paper]{article}
\usepackage{ifthen} % for conditional statements
\newboolean{pdflatex}
\setboolean{pdflatex}{true} % False for eps figures

\newboolean{articletitles}
\setboolean{articletitles}{true} % False removes titles in references

\newboolean{uprightparticles}
\setboolean{uprightparticles}{false} %True for upright particle symbols

\pdfoutput=1
\newboolean{inbibliography}
\setboolean{inbibliography}{false} %True once you enter the bibliography

\usepackage{microtype}
\usepackage{lineno}  % for line numbering during review
\usepackage{xspace} % To avoid problems with missing or double spaces after
\usepackage{graphicx}  % to include figures (can also use other packages)
\usepackage{color}
\usepackage{colortbl}
\graphicspath{{./figs/}} % Make Latex search fig subdir for figures

\usepackage{amsmath} % Adds a large collection of math symbols
\usepackage{amssymb}
\usepackage{amsfonts}
\usepackage{upgreek} % Adds in support for greek letters in roman typeset

\newcommand*\patchAmsMathEnvironmentForLineno[1]{%
\expandafter\let\csname old#1\expandafter\endcsname\csname #1\endcsname
\expandafter\let\csname oldend#1\expandafter\endcsname\csname
end#1\endcsname
 \renewenvironment{#1}%
   {\linenomath\csname old#1\endcsname}%
   {\csname oldend#1\endcsname\endlinenomath}%
}
\newcommand*\patchBothAmsMathEnvironmentsForLineno[1]{%
  \patchAmsMathEnvironmentForLineno{#1}%
  \patchAmsMathEnvironmentForLineno{#1*}%
}
\AtBeginDocument{%
\patchBothAmsMathEnvironmentsForLineno{equation}%
\patchBothAmsMathEnvironmentsForLineno{align}%
\patchBothAmsMathEnvironmentsForLineno{flalign}%
\patchBothAmsMathEnvironmentsForLineno{alignat}%
\patchBothAmsMathEnvironmentsForLineno{gather}%
\patchBothAmsMathEnvironmentsForLineno{multline}%
}

\usepackage{hyperref}    % Hyperlinks in references
\usepackage[all]{hypcap} % Internal hyperlinks to floats.

\def\lhcb {\mbox{LHCb}\xspace}

\ifthenelse{\boolean{uprightparticles}}%
{

 \def\Peta        {\ensuremath{\upeta}\xspace}

 \def\Ppi         {\ensuremath{\uppi}\xspace}

 \def\Ppsi        {\ensuremath{\uppsi}\xspace}

 \def\PDelta      {\ensuremath{\Delta}\xspace}                 
 \def\PXi      {\ensuremath{\Xi}\xspace}                 
 \def\PLambda      {\ensuremath{\Lambda}\xspace}                 
 \def\PSigma      {\ensuremath{\Sigma}\xspace}                 
 \def\POmega      {\ensuremath{\Omega}\xspace}                 
 \def\PUpsilon      {\ensuremath{\Upsilon}\xspace}                 
 
 \def\PB      {\ensuremath{\mathrm{B}}\xspace}                 
                  
 \def\PD      {\ensuremath{\mathrm{D}}\xspace}

 \def\PJ      {\ensuremath{\mathrm{J}}\xspace}                 
 \def\PK      {\ensuremath{\mathrm{K}}\xspace}

 \def\Pc      {\ensuremath{\mathrm{c}}\xspace}                 
 \def\Pd      {\ensuremath{\mathrm{d}}\xspace}

 \def\Pi      {\ensuremath{\mathrm{i}}\xspace}

 \def\Pp      {\ensuremath{\mathrm{p}}\xspace}

 \def\Ps      {\ensuremath{\mathrm{s}}\xspace}                 
                  
 \def\Pu      {\ensuremath{\mathrm{u}}\xspace}

}
{

 \def\Peta        {\ensuremath{\eta}\xspace}

 \def\Ppi         {\ensuremath{\pi}\xspace}

 \def\Ppsi        {\ensuremath{\psi}\xspace}                 
                  
 \mathchardef\PDelta="7101
 \mathchardef\PXi="7104
 \mathchardef\PLambda="7103
 \mathchardef\PSigma="7106
 \mathchardef\POmega="710A
 \mathchardef\PUpsilon="7107
                  
 \def\PB      {\ensuremath{B}\xspace}                 
                  
 \def\PD      {\ensuremath{D}\xspace}

 \def\PJ      {\ensuremath{J}\xspace}                 
 \def\PK      {\ensuremath{K}\xspace}

 \def\Pc      {\ensuremath{c}\xspace}                 
 \def\Pd      {\ensuremath{d}\xspace}

 \def\Pi      {\ensuremath{i}\xspace}

 \def\Pp      {\ensuremath{p}\xspace}

 \def\Ps      {\ensuremath{s}\xspace}                 
                  
 \def\Pu      {\ensuremath{u}\xspace}

}

\def\uquark    {\ensuremath{\Pu}\xspace}
\def\uquarkbar {\ensuremath{\overline \uquark}\xspace}

\def\dquark    {\ensuremath{\Pd}\xspace}
\def\dquarkbar {\ensuremath{\overline \dquark}\xspace}

\def\squark    {\ensuremath{\Ps}\xspace}
\def\squarkbar {\ensuremath{\overline \squark}\xspace}

\def\cquark    {\ensuremath{\Pc}\xspace}

\def\pion  {\ensuremath{\Ppi}\xspace}

\def\pip   {\ensuremath{\pion^+}\xspace}
\def\pim   {\ensuremath{\pion^-}\xspace}

\def\kaon  {\ensuremath{\PK}\xspace}
  \def\Kbar  {\kern 0.2em\overline{\kern -0.2em \PK}{}\xspace}

\def\Kp    {\ensuremath{\kaon^+}\xspace}
\def\Km    {\ensuremath{\kaon^-}\xspace}
\def\Kpm   {\ensuremath{\kaon^\pm}\xspace}

  \def\Dbar    {\kern 0.2em\overline{\kern -0.2em \PD}{}\xspace}

\def\B       {\ensuremath{\PB}\xspace}
\def\Bbar    {\ensuremath{\kern 0.18em\overline{\kern -0.18em \PB}{}}\xspace}

\def\Bu      {\ensuremath{\B^+}\xspace}

\def\Bp      {\ensuremath{\Bu}\xspace}

\def\Bpm     {\ensuremath{\B^\pm}\xspace}

\def\jpsi     {\ensuremath{{\PJ\mskip -3mu/\mskip -2mu\Ppsi\mskip 2mu}}\xspace}
\def\psitwos  {\ensuremath{\Ppsi{(2S)}}\xspace}

\def\etac     {\ensuremath{\Peta_\cquark}\xspace}

  \def\Y#1S{\ensuremath{\PUpsilon{(#1S)}}\xspace}% no space before {...}!

\def\proton      {\ensuremath{\Pp}\xspace}
\def\antiproton  {\ensuremath{\overline \proton}\xspace}

\def\Lz {\ensuremath{\PLambda}\xspace}
\def\Lbar {\ensuremath{\kern 0.1em\overline{\kern -0.1em\PLambda}}\xspace}

\def\Sigmares {\ensuremath{\PSigma}\xspace}
\def\Sigmaresbar{\ensuremath{\overline \Sigmares}\xspace}

\def\to                 {\ensuremath{\rightarrow}\xspace}

\def\CP                {\ensuremath{C\!P}\xspace}
\def\CPT               {\ensuremath{C\!PT}\xspace}

\def\AFB      {\ensuremath{A_{\mathrm{FB}}}\xspace}

\def\AT#1     {\ensuremath{A_{\mathrm{T}}^{#1}}\xspace}           % 2

\def\C#1      {\ensuremath{\mathcal{C}_{#1}}\xspace}                       % 9
\def\Cp#1     {\ensuremath{\mathcal{C}_{#1}^{'}}\xspace}                    % 7
\def\Ceff#1   {\ensuremath{\mathcal{C}_{#1}^{\mathrm{(eff)}}}\xspace}        % 9  
\def\Cpeff#1  {\ensuremath{\mathcal{C}_{#1}^{'\mathrm{(eff)}}}\xspace}       % 7
\def\Ope#1    {\ensuremath{\mathcal{O}_{#1}}\xspace}                       % 2
\def\Opep#1   {\ensuremath{\mathcal{O}_{#1}^{'}}\xspace}                    % 7

\newcommand{\tev}{\ifthenelse{\boolean{inbibliography}}{\ensuremath{~T\kern -0.05em eV}\xspace}{\ensuremath{\mathrm{\,Te\kern -0.1em V}}\xspace}}
\newcommand{\gev}{\ensuremath{\mathrm{\,Ge\kern -0.1em V}}\xspace}
\newcommand{\mev}{\ensuremath{\mathrm{\,Me\kern -0.1em V}}\xspace}
\newcommand{\kev}{\ensuremath{\mathrm{\,ke\kern -0.1em V}}\xspace}
\newcommand{\ev}{\ensuremath{\mathrm{\,e\kern -0.1em V}}\xspace}
\newcommand{\gevc}{\ensuremath{{\mathrm{\,Ge\kern -0.1em V\!/}c}}\xspace}
\newcommand{\mevc}{\ensuremath{{\mathrm{\,Me\kern -0.1em V\!/}c}}\xspace}
\newcommand{\gevcc}{\ensuremath{{\mathrm{\,Ge\kern -0.1em V\!/}c^2}}\xspace}
\newcommand{\gevccbis}{\ensuremath{{\mathrm{Ge\kern -0.1em V\!/}c^2}}\xspace}
\newcommand{\gevgevcccc}{\ensuremath{{\mathrm{\,Ge\kern -0.1em V^2\!/}c^4}}\xspace}
\newcommand{\mevcc}{\ensuremath{{\mathrm{\,Me\kern -0.1em V\!/}c^2}}\xspace}

\def\mum  {\ensuremath{\,\upmu\rm m}\xspace}

\def\invfb   {\ensuremath{\mbox{\,fb}^{-1}}\xspace}

\newcommand{\chisq}{\ensuremath{\chi^2}\xspace}

\def\gsim{{~\raise.15em\hbox{$>$}\kern-.85em
          \lower.35em\hbox{$\sim$}~}\xspace}
\def\lsim{{~\raise.15em\hbox{$<$}\kern-.85em
          \lower.35em\hbox{$\sim$}~}\xspace}

\def\pt         {\mbox{$p_{\rm T}$}\xspace}

\def\evtgen     {\mbox{\textsc{EvtGen}}\xspace}

\def\geant      {\mbox{\textsc{Geant4}}\xspace}

\def\photos     {\mbox{\textsc{Photos}}\xspace}

\def\pythia     {\mbox{\textsc{Pythia}}\xspace}

\def\tell1  {TELL1\xspace}
\def\ukl1   {UKL1\xspace}

\usepackage{cite} % Allows for ranges in citations
\usepackage{mciteplus}
\usepackage[numbers, square, comma, sort&compress]{natbib}
\usepackage[bottom]{footmisc}
\usepackage{booktabs}
\usepackage{dcolumn}

\usepackage{longtable} % only for template; not usually to be used in PAPERs

\def\pppi {\ensuremath{\Bpm \to p \antiproton \pi^{\pm}}\xspace}
\def\ppk {\ensuremath{\Bpm \to p \antiproton K^{\pm}}\xspace}

\def\hhhplus {\ensuremath{B^+ \to h^+ h^+ h^-}\xspace}
\def\pipipiplus {\ensuremath{B^+ \to \pip \pip \pim}\xspace}
\def\kpipiplus {\ensuremath{B^+ \to \Kp \pip \pim}\xspace}
\def\kkpiplus {\ensuremath{B^+ \to \Kp \Km \pip}\xspace}
\def\kkkplus {\ensuremath{B^+ \to \Kp \Kp \Km}\xspace}
\def\pppiplus {\ensuremath{B^+ \to p \antiproton  \pip}\xspace}
\def\ppkplus {\ensuremath{B^+ \to p \antiproton  \Kp}\xspace}
\def\pphplus {\ensuremath{B^+ \to p \antiproton  h^+}\xspace}

\def\lambdaFifteenTwentypplus {\ensuremath{B^+ \to \Lbar(1520)p}\xspace}
\def\acp {\ensuremath{A_{\CP}}\xspace}

\def\acpraw {\ensuremath{A_{\rm raw}}\xspace}

\def\aprod {\ensuremath{A_{\rm P}}\xspace}
\def\adet {\ensuremath{A_{\rm D}}\xspace}

\begin{document}

%%%%%%%%%%%%%%%%%%%%%%%%%
%%%%% Title     %%%%%%%%%
%%%%%%%%%%%%%%%%%%%%%%%%%
\renewcommand{\thefootnote}{\fnsymbol{footnote}}
\setcounter{footnote}{1}

% %%%%%%% CHOOSE TITLE PAGE--------
%\onecolumn
% \input{title-LHCb-ANA}
%\input{title-LHCb-CONF}
% $Id: title-LHCb-PAPER.tex 33814 2013-04-13 10:23:21Z roldeman $
% ===============================================================================
% Purpose: LHCb-PAPER journal paper title page template
% Author:
% Created on: 2010-09-25
% ===============================================================================

%%%%%%%%%%%%%%%%%%%%%%%%%
%%%%%  TITLE PAGE  %%%%%%
%%%%%%%%%%%%%%%%%%%%%%%%%
\begin{titlepage}
\pagenumbering{roman}

% Header ---------------------------------------------------
\vspace*{-1.5cm}
\centerline{\large EUROPEAN ORGANIZATION FOR NUCLEAR RESEARCH (CERN)}
\vspace*{1.5cm}
\hspace*{-0.5cm}
\begin{tabular*}{\linewidth}{lc@{\extracolsep{\fill}}r}
\ifthenelse{\boolean{pdflatex}}% Logo format choice
{\vspace*{-2.7cm}\mbox{\!\!\!\includegraphics[width=.14\textwidth]{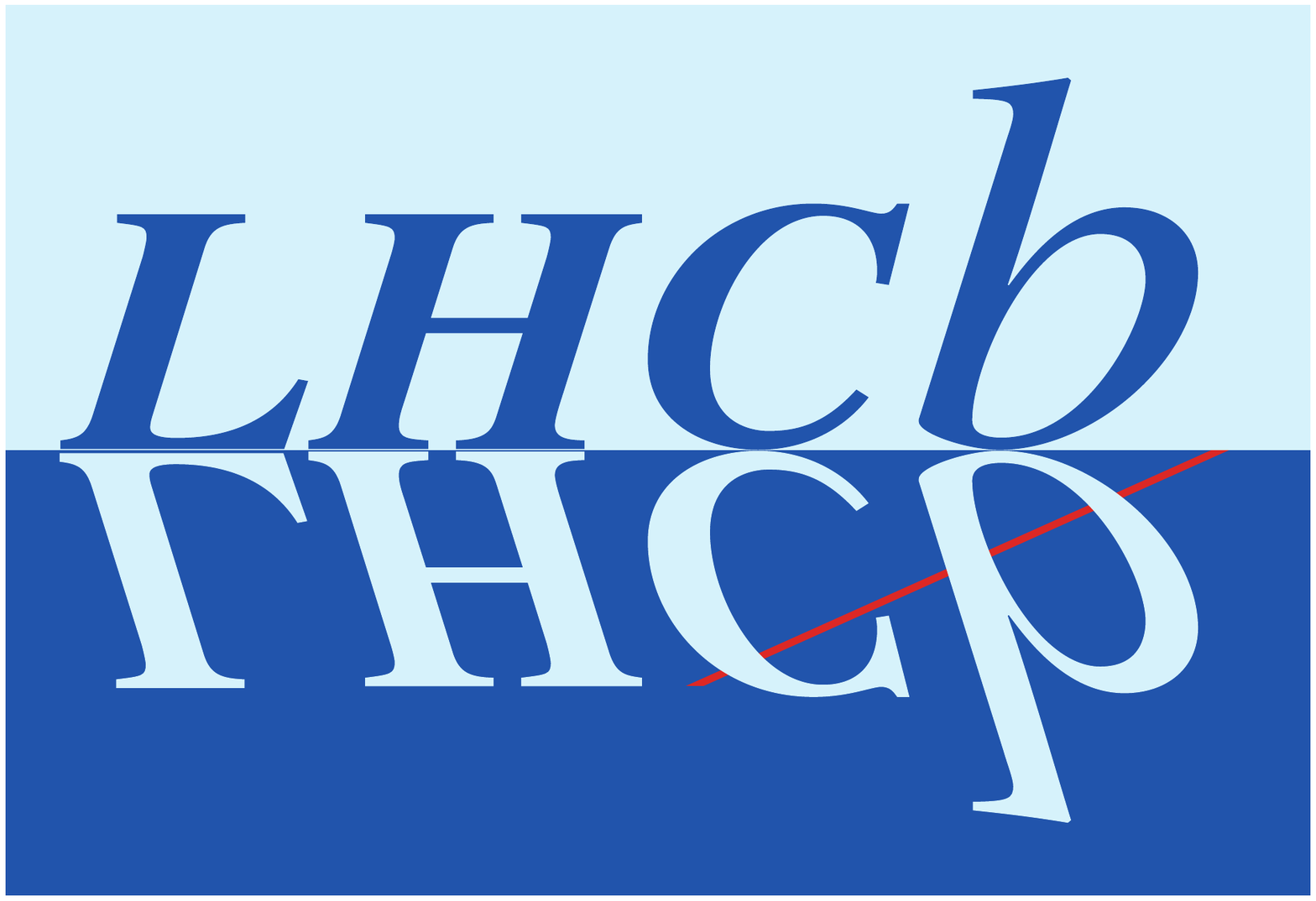}} & &}%
{\vspace*{-1.2cm}\mbox{\!\!\!\includegraphics[width=.12\textwidth]{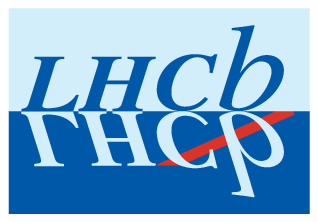}} & &}%
\\
 & & CERN-PH-EP-2013-125 \\  % ID
 & & LHCb-PAPER-2013-031 \\  % ID
 & & July, 23 2013 \\ % Date - Can also hardwire e.g.: 23 March 2010
% & & Version 2.4\\
% not in paper \hline
\end{tabular*}

%\vspace*{4.0cm}

% Title --------------------------------------------------
{\bf\boldmath\large
\begin{center}
Studies of the decays \pphplus and observation of \lambdaFifteenTwentypplus 
\end{center}
}

\vspace*{2.0cm}
\begin{center}
The LHCb collaboration
\end{center}

% Authors -------------------------------------------------

%\input{LHCb_HD_authorlist_2013-06-03.tex}

%\vspace{\fill}

% Abstract -----------------------------------------------

\begin{abstract}
  \noindent
Dynamics and direct \CP violation in three-body charmless decays of charged $B$ mesons to a proton, an antiproton and a light meson (pion or kaon) are studied using data, corresponding to an integrated luminosity of 1.0\invfb, collected by the \lhcb experiment in $pp$ collisions at a center-of-mass energy of $7$ TeV. Production spectra are determined as a function of Dalitz-plot and helicity variables. The forward-backward asymmetry of the light meson in the $p\antiproton$ rest frame is measured. No significant \CP asymmetry in \ppkplus decay is found in any region of the Dalitz plane. We present the first observation of the decay $B^+\to\Lbar(1520)(\to K^+\antiproton)p$ near the $K^+\antiproton$ threshold and measure $\mathcal{B}(\lambdaFifteenTwentypplus)=(3.9^{+1.0}_{-0.9}~(\mathrm{stat})\pm0.1~(\mathrm{syst})\pm0.3~(\mathrm{BF}))\times 10^{-7}$, where BF denotes the uncertainty on secondary branching fractions.
\end{abstract} 

\vspace*{2.0cm}

\begin{center}
  Submitted to Phys.~Rev.~D
\end{center}
%\begin{center}
%  Submitted to JHEP / Phys.~Rev.~D / Phys.~Rev.~Lett. / Phys.~Lett.~B / Eur.~Phys.~J.~C / Nucl.~Phys.~B
%\end{center}

%\vspace{\fill}

{\footnotesize
\centerline{\copyright~CERN on behalf of the \lhcb collaboration, license \href{http://creativecommons.org/licenses/by/3.0/}{CC-BY-3.0}.}}
\vspace*{2mm}

\end{titlepage}

%%%%%%%%%%%%%%%%%%%%%%%%%%%%%%%%
%%%%%  EOD OF TITLE PAGE  %%%%%%
%%%%%%%%%%%%%%%%%%%%%%%%%%%%%%%%

%  empty page follows the title page ----
%\newpage
%\setcounter{page}{2}
%\mbox{~}
%\newpage

% Author List ----------------------------
%  You need to get a new author list!

%%%%%%%%%%%%%%%%%%%%%%%%%%%%%%%%%%%%%%%%%%
\centerline{\large\bf LHCb collaboration}
\begin{flushleft}
\small
R.~Aaij$^{40}$, 
B.~Adeva$^{36}$, 
M.~Adinolfi$^{45}$, 
C.~Adrover$^{6}$, 
A.~Affolder$^{51}$, 
Z.~Ajaltouni$^{5}$, 
J.~Albrecht$^{9}$, 
F.~Alessio$^{37}$, 
M.~Alexander$^{50}$, 
S.~Ali$^{40}$
% ...

{\footnotesize \it
% ...
$ ^{5}$Clermont Universit\'{e}, Universit\'{e} Blaise Pascal, CNRS/IN2P3, LPC, Clermont-Ferrand, France\\
$ ^{6}$CPPM, Aix-Marseille Universit\'{e}, CNRS/IN2P3, Marseille, France\\
% ...
$ ^{9}$Fakult\"{a}t Physik, Technische Universit\"{a}t Dortmund, Dortmund, Germany\\
% ...
$ ^{36}$Universidad de Santiago de Compostela, Santiago de Compostela, Spain\\
$ ^{37}$European Organization for Nuclear Research (CERN), Geneva, Switzerland\\
% ...
$ ^{40}$Nikhef National Institute for Subatomic Physics, Amsterdam, The Netherlands\\
% ...
$ ^{45}$H.H. Wills Physics Laboratory, University of Bristol, Bristol, United Kingdom\\
% ...
$ ^{50}$School of Physics and Astronomy, University of Glasgow, Glasgow, United Kingdom\\
$ ^{51}$Oliver Lodge Laboratory, University of Liverpool, Liverpool, United Kingdom\\
% ...
}
\end{flushleft}
%%%%%%%%%%%%%%%%%%%%%%%%%%%%%%%%%%%%%%%%%%

\cleardoublepage
%\clearpage

%\twocolumn
% %%%%%%%%%%%%% ---------

\renewcommand{\thefootnote}{\arabic{footnote}}
\setcounter{footnote}{0}

%%%%%%%%%%%%%%%%%%%%%%%%%%%%%%%%
%%%%%  Table of Content   %%%%%%
%%%%%%%%%%%%%%%%%%%%%%%%%%%%%%%%
%%%% Uncomment next 2 lines if desired
%\tableofcontents
%\cleardoublepage

%%%%%%%%%%%%%%%%%%%%%%%%%
%%%%% Main text %%%%%%%%%
%%%%%%%%%%%%%%%%%%%%%%%%%

\pagestyle{plain} % restore page numbers for the main text
\setcounter{page}{1}
\pagenumbering{arabic}

%% Uncomment during review phase.
%% Comment before a final submission.
%\linenumbers

% You can include short sections directly in the main tex file.
% However, for larger papers it is desirable to split the text into
% several semiautonomous files, which can be revised independently.
% This is especially useful when developing a document in
% collaboration with several people, since then different parts can be
% edited independently.  This type of file organization is shown here.
%

\section{Introduction}
Evidence of inclusive direct \CP violation in three-body charmless decays of $B^+$ mesons\footnote{Throughout the paper, the inclusion of charge conjugate processes is implied, except in the definition of \CP asymmetries.} has recently been found in the modes \kpipiplus, \kkkplus, \pipipiplus, and \kkpiplus~\cite{LHCb-CONF-2012-028,LHCb-PAPER-2013-027}. In addition, very large \CP asymmetries were observed in the low $K^+K^-$ and $\pi^+\pi^-$ mass regions, without clear connection to a resonance. The localization of the asymmetries and the correlation of the \CP violation between the decays suggest that $\pi^+\pi^- \leftrightarrow K^+K^-$ rescattering may play an important role in the generation of the strong phase difference needed for such a violation to occur \cite{Riazuddin,Wolfenstein}. Conservation of \CPT symmetry imposes a constraint on the sum of the rates of final states with the same flavour quantum numbers, providing the possibility of entangled long-range effects contributing to the \CP violating mechanism \cite{Compound}. In contrast, $h^+h^-\leftrightarrow p\antiproton$ ($h=\pi$ or $K$ throughout the paper) rescattering is expected to be suppressed compared to $\pi^+\pi^- \leftrightarrow K^+K^-$, and thus is not expected to play an important role.

\begin{figure}[b]%bt\centering
\begin{center}
\includegraphics[width=0.5\textwidth]{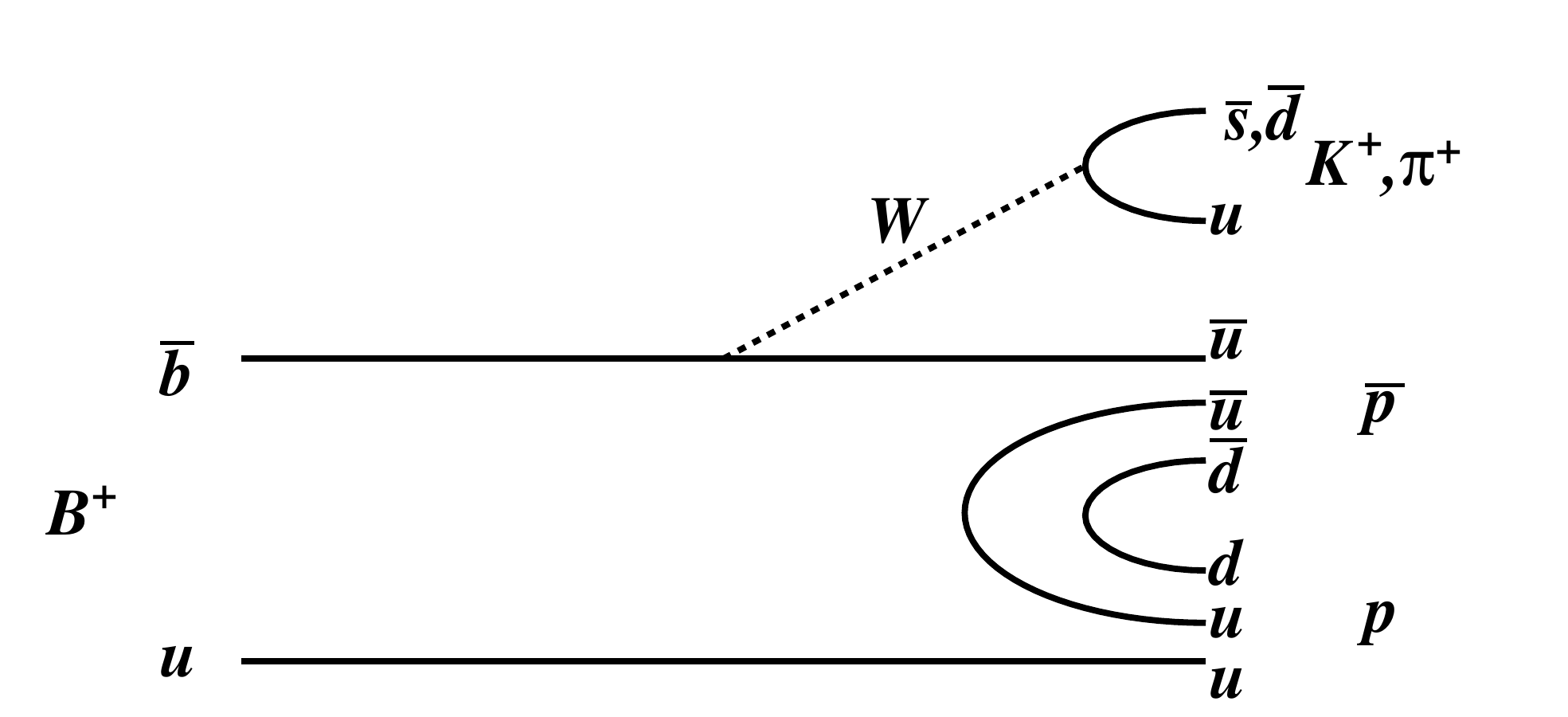}%[scale=.45]
\includegraphics[width=0.5\textwidth]{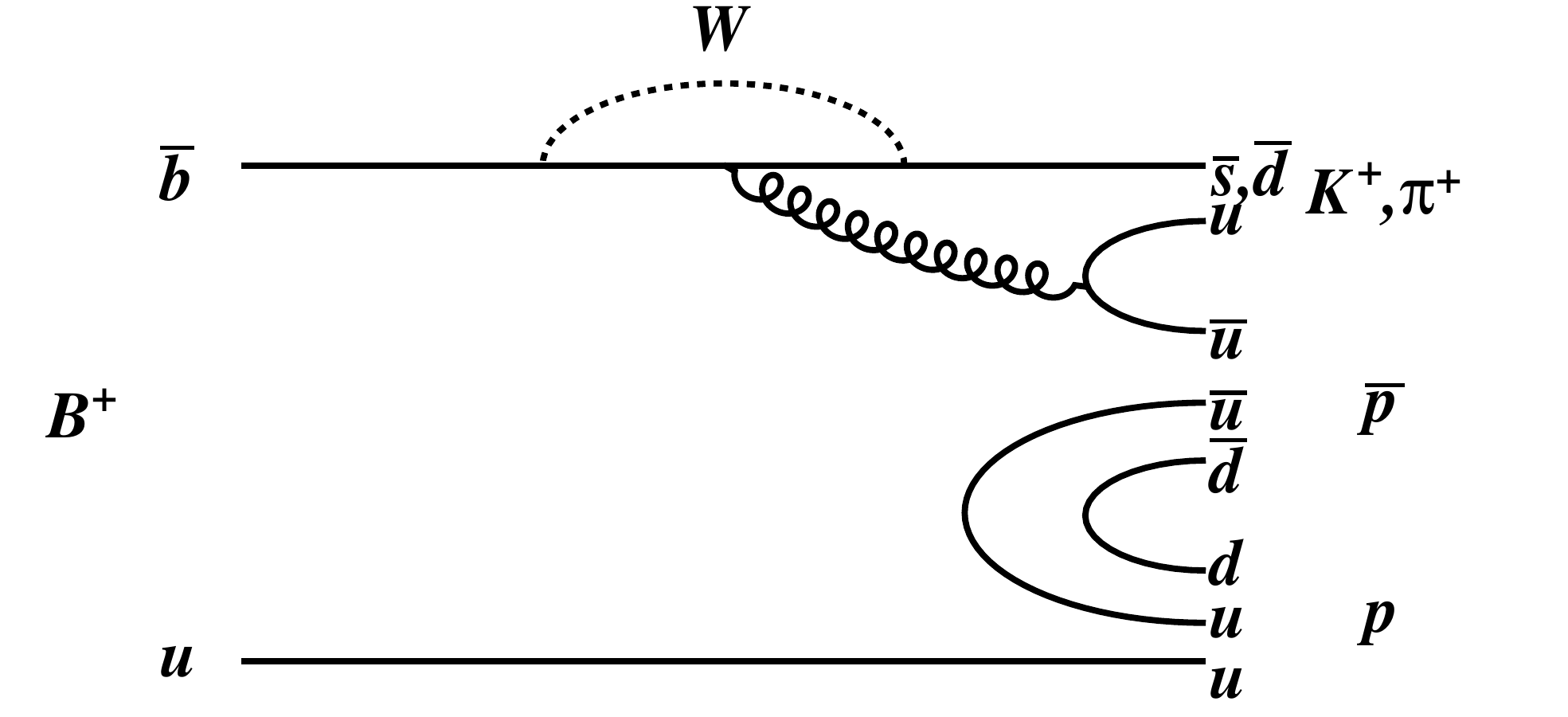}%[scale=.45]
\end{center}
\caption{Leading tree and penguin diagrams for \pphplus decays.}
\label{Fig:feyn_dia}
\end{figure}

The leading quark-level diagrams for the modes \pphplus are shown in Fig.~\ref{Fig:feyn_dia}. The \ppkplus mode is expected to be dominated by the $b\to s$ loop (penguin) transition while the mode \pppiplus is likely to be dominated by the $b\to u$ tree decay, which is CKM suppressed compared to the former.
Since the short distance dynamics are similar to that of the \hhhplus modes, a \CP analysis of \pphplus decays could help to clarify the role of long-range scatterings in the \CP asymmetries of \hhhplus decays.

First studies were performed at the $B$ factories on the production and dynamics of \pphplus decays \cite{BaBar_pph1,BaBar_pph2,Belle_pph}. The results have shown a puzzling opposite behaviour of \ppkplus and \pppiplus decays in the asymmetric occupation of the Dalitz plane. Charmonium contributions to the \ppkplus decay have been studied by \lhcb \cite{LHCb-PAPER-2012-047}.
This paper reports a detailed study of the dynamics of the \pphplus decays and a systematic search for \CP violation, both inclusively and in regions of the Dalitz plane. The charmless region, defined for the invariant mass $m_{p\antiproton}<2.85\gevcc$, is of particular interest. The relevant observables are the differential production spectra of Dalitz-plot variables and the global charge asymmetry \acp, defined as
\begin{equation}
\acp = \frac{N(B^-\to f^-)-N(B^+\to f^+)}{N(B^-\to f^-)+N(B^+\to f^+)},
\end{equation}
where $f^{\pm}=p\antiproton h^{\pm}$. The mode $\Bp \to \jpsi(\to p\antiproton)\Kp$ serves as a control channel. The first observation of the decay \lambdaFifteenTwentypplus is presented. Its branching fraction is derived through the ratio of its yield to the measured yield of the $\Bp \to \jpsi(\to p\antiproton)\Kp$ decay.

\section{Detector and software}
\label{sec:Detector}

The \lhcb detector~\cite{Alves:2008td} is a single-arm forward
spectrometer covering the \mbox{pseudorapidity} range $2<\eta <5$,
designed for the study of particles containing \b or \c quarks. The detector includes a high precision tracking system
consisting of a silicon-strip vertex detector surrounding the $pp$
interaction region, a large-area silicon-strip detector located
upstream of a dipole magnet with a bending power of about
$4{\rm\,Tm}$, and three stations of silicon-strip detectors and straw
drift tubes placed downstream. The combined tracking system has momentum resolution $\Delta p/p$ that varies from 0.4\% at 5\gevc to
0.6\% at 100\gevc, and impact parameter~(IP) resolution of 20\mum for
tracks with high transverse momentum. Charged hadrons are identified
using two ring-imaging Cherenkov detectors (RICH) \cite{RichPerf}. Photon, electron and
hadron candidates are identified by a calorimeter system consisting of
scintillating-pad and preshower detectors, an electromagnetic
calorimeter and a hadronic calorimeter. Muons are identified by a
system composed of alternating layers of iron and multiwire
proportional chambers. 

The trigger~\cite{Aaij:2012me} consists of a hardware stage, based
on information from the calorimeter and muon systems, followed by a
software stage which applies a full event reconstruction. Events triggered both on objects independent of the signal, and associated with the signal, are used. In the latter case, the transverse energy of the hadronic cluster is required to be at least 3.5\gev.
The software trigger requires a two-, three- or four-track secondary vertex with a large sum of the transverse momentum, \pt, of the tracks and a significant displacement from all primary $pp$ interaction vertices.  At least one track must have $\pt > 1.7\gevc$, track fit \chisq per degree of freedom less than 2, and an impact parameter \chisq ($\chisq_{\mathrm{IP}}$) with respect to any primary interaction greater than 16. The $\chisq_{\mathrm{IP}}$ is defined as the difference between the \chisq of the primary vertex reconstructed with and without the considered track. A multivariate algorithm is used to identify secondary vertices~\cite{Gligorov:2012qt}.

The simulated $pp$ collisions are generated using
\pythia~6.4~\cite{Sjostrand:2006za} with a specific \lhcb
configuration~\cite{LHCb-PROC-2010-056}.  Decays of hadronic particles
are described by \evtgen~\cite{Lange:2001uf} in which final state
radiation is generated using \photos~\cite{Golonka:2005pn}. The
interaction of the generated particles with the detector and its
response are implemented using the \geant
toolkit~\cite{Allison:2006ve,*Agostinelli:2002hh} as described in
Ref.~\cite{LHCb-PROC-2011-006}. Non-resonant \pphplus events are simulated, uniformly distributed in phase space, to study the variation of efficiencies across the Dalitz plane, as well as resonant samples such as $\Bp \to \jpsi(\to p\antiproton)\Kp$, $\Bp \to \etac(\to p\antiproton)\Kp$, $\Bp \to \psitwos(\to p\antiproton)\Kp$, $\Bp \to \Lbar(1520)(\to K^+\antiproton)p$, and $\Bp \to \jpsi(\to p\antiproton)\pip$.

\section{Signal reconstruction and determination}
Candidate \pphplus decays are formed by combining three charged tracks, with appropriate mass assignments. The tracks are required to satisfy track fit quality criteria and a set of loose selection requirements on their momenta, transverse momenta, $\chisq_{\mathrm{IP}}$, and distance of closest approach between any pair of tracks. The requirement on the momentum of the proton candidates, $p>3\gevc$, is larger than for the kaon and pion candidates, $p>1.5\gevc$. The $\Bp$ candidates formed by the combinations are required to have $p_{\mathrm{T}}>$ 1.7~\gevc and $\chisq_{\mathrm{IP}}<10$. The distance between the decay vertex and the primary vertex is required to be greater than 3 mm, and the vector formed by the primary and decay vertices must align with the $\Bp$ candidate momentum. Particle identification (PID) is applied to the proton, kaon and pion candidates, using combined subdetector information, the main separation power being provided by the RICH system. The PID efficiencies are derived from data calibration samples of kinematically identified pions, kaons and protons originating from the decays $D^{*+}\to D^0(\to K^-\pi^+)\pi^+$ and $\Lz\to p\pi^-$.

Signal and background are extracted using unbinned extended maximum likelihood fits to the mass of the $p\antiproton h^+$ combinations.
The \ppkplus signal is modelled by a double Gaussian function. The combinatorial background is represented by a second-order polynomial function. A Gaussian function accounting for a partially reconstructed component from $B\to p\antiproton K^*$ decays is used. A possible $p\antiproton \pip$ cross-feed contribution is included in the fit and is found to be small. An asymmetric Gaussian function with power law tails is used to estimate the uncertainties related to the variation of the signal yield.

In the case of the \pppiplus decay, the signal yield is smaller and the background is larger. The ranges of the signal and cross-feed parameters are constrained to the values obtained in the simulation within their uncertainties. The signal and the $p\antiproton \Kp$ cross-feed contribution are modelled with Gaussian functions. The combinatorial background is represented by a third-order polynomial function.

\begin{figure}[t]%bt\centering
\begin{center}
\includegraphics[width=0.45\textwidth]{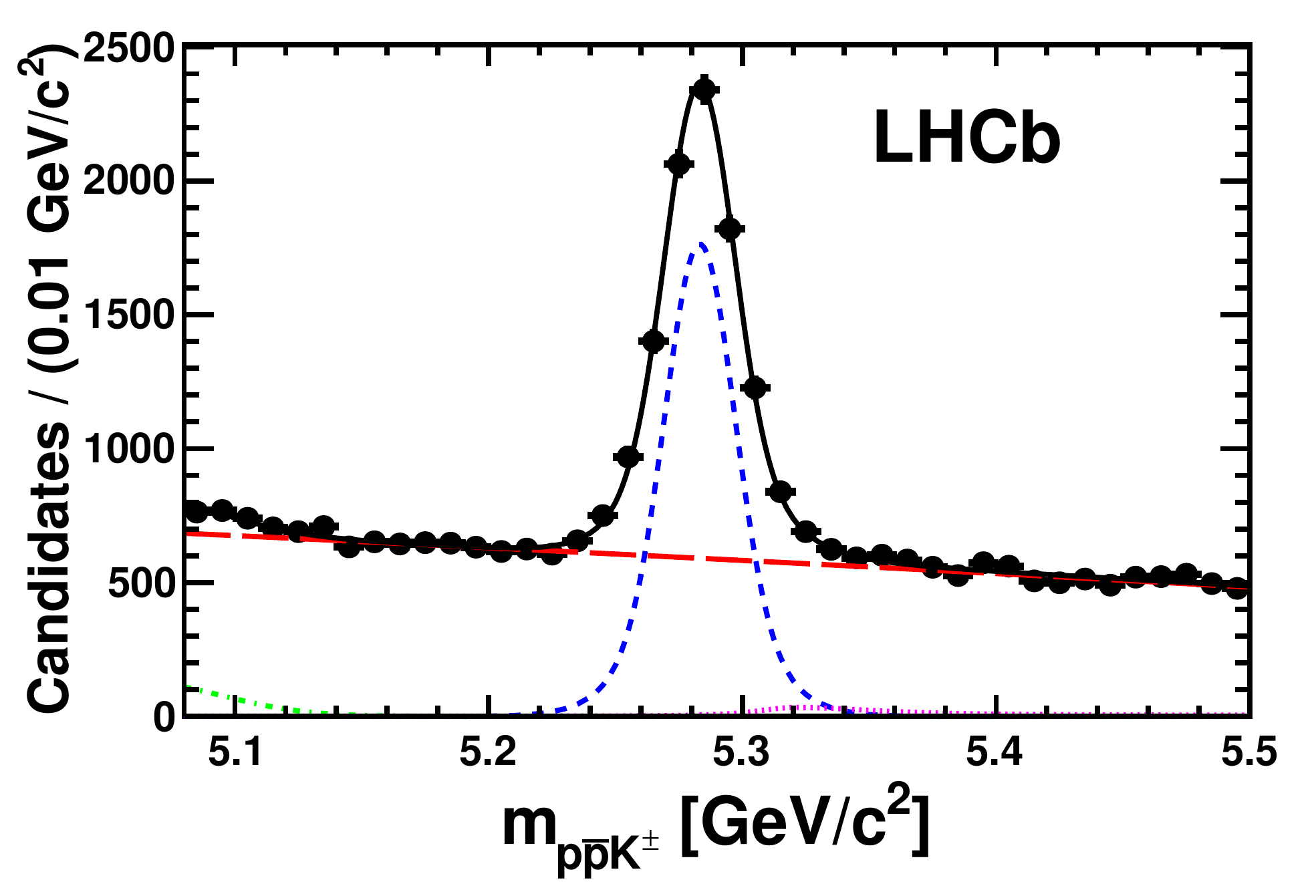}
\includegraphics[width=0.45\textwidth]{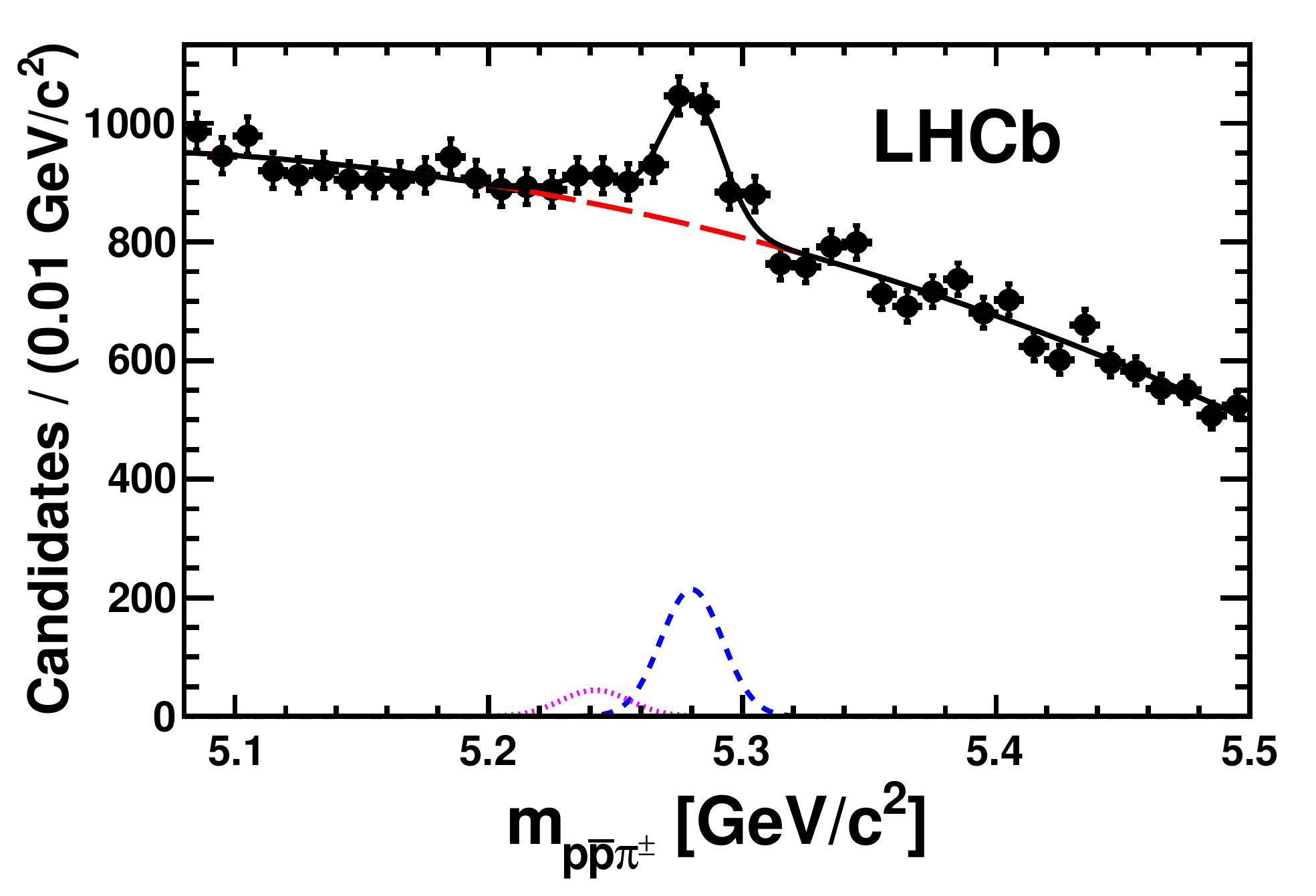}
\end{center}
\caption{Invariant mass distributions of (left) $p\antiproton \Kp$ and (right) $p\antiproton \pip$ candidates. The points with error bars represent data. The solid black line represents the total fit function. Blue dashed, purple dotted, red long-dashed and green dashed-dotted curves represent the signal, cross-feed, combinatorial background and partially reconstructed background, respectively.}
\label{Fig:pph_globalfits}
\end{figure}

The \pphplus invariant mass spectra are shown in Fig.~\ref{Fig:pph_globalfits}.
%\onecolumn
%\twocolumn
The signal yields obtained from the fits are $N(p\antiproton K^\pm)=7029\pm139$ and $N(p\antiproton \pi^\pm)=656\pm70$, where the uncertainties are statistical only.

\section{Dynamics of \boldmath\pphplus decays}
To probe the dynamics of the \pphplus decays, differential production spectra are derived as a function of $m_{p\antiproton }$ and $\cos\theta_p$, where $\theta_p$ is the angle between the charged meson $h$ and the opposite-sign baryon in the rest frame of the $p\antiproton $ system. The $p\antiproton h^+$ invariant mass is fitted in bins of the aforementioned variables and the signal yields are corrected for trigger, reconstruction and selection efficiencies. They are estimated with simulated samples and corrected to account for discrepancies between data and simulation. The signal yields are determined with the fit models described in the previous section, but allowing the combinatorial background parameters to vary. The systematic uncertainties are determined for each bin and include uncertainties related to the PID correction, fit model, trigger efficiency, and the size of the simulated samples. The latter is evaluated from the differences between data and simulation as a function of the Dalitz-plot variables. No trigger-induced distortions are found.

\subsection{Invariant mass of the \boldmath$p\antiproton $ system}

\begin{table}[tb]
\caption{ Fitted \ppkplus signal yield, including the charmonium modes, efficiency and relative systematic uncertainty, in bins of $p\antiproton $ invariant mass. The error on the efficiency includes all the sources of uncertainty.}
  \begin{center}
\begin{tabular}{l r@{$\pm$}l c c}\hline\hline
\centering
 $m_{p\antiproton }$ $[\gevcc]$ & \multicolumn{2}{c}{\ppkplus yield} & \multicolumn{1}{c}{Efficiency (\%)} & \multicolumn{1}{c}{Syst. (\%)}
\\
\hline 
$<$ 2.85 & \hspace{5mm}3315&83 & 1.74$\pm$0.04&2.9\\
\hline
$<$ 2 &\hspace{5mm}446&32 &1.80$\pm$0.08 & 8.1\\
$[2,2.2]$ &\hspace{5mm}1001&42 &1.77$\pm$0.05 & 4.4\\
$[2.2,2.4]$ &\hspace{5mm}732&39 &1.77$\pm$0.03 & 4.0\\
$[2.4,2.6]$ &\hspace{5mm}550&35 &1.67$\pm$0.03 & 3.4\\
$[2.6,2.85]$ &\hspace{5mm}580&34 &1.67$\pm$0.02 & 2.9\\
\hline
$[2.85,3.15]$ & \hspace{5mm}2768&58 &1.61$\pm$0.02 & 2.6\\
$[3.15,3.3]$ & \hspace{5mm}125&18 &1.57$\pm$0.03 & 3.8\\
$[3.3,4]$ & \hspace{5mm}585&37 &1.47$\pm$0.01 & 2.2\\
$>$ 4 & \hspace{5mm}233&32 &1.22$\pm$0.01 & 2.3\\
\hline\hline
\end{tabular}
\end{center}
    \label{Tab:ppK_mppbins}
\end{table}

\begin{table}[tb]
\caption{ Fitted \pppiplus signal yield, including the \jpsi mode, \ppkplus cross-feed yield, signal efficiency, and relative systematic uncertainty in bins of $p\antiproton $ invariant mass.}
  \begin{center}
\begin{tabular}{l r@{$\pm$}l r@{$\pm$}l c c}\hline\hline
\centering
 $m_{p\antiproton }$ $[\gevcc]$ & \multicolumn{2}{c}{\pppiplus yield} & \multicolumn{2}{c}{\ppkplus cross-feed} & \multicolumn{1}{c}{Efficiency (\%)} & \multicolumn{1}{c}{Syst. (\%)}\\ 
\hline 
$<$ 2.85 & \hspace{5mm}564&61& \hspace{10mm}114&62&1.31$\pm$0.10 & 7.6\\
\hline
$<$ 2 & \hspace{5mm}140&26 & \hspace{10mm}64&26 &1.34$\pm$0.15 & 11\\
$[2,2.2]$ & \hspace{5mm}261&31 & \hspace{10mm}10&29 &1.30$\pm$0.10 &7.9\\
$[2.2,2.4]$ & \hspace{5mm}95&30 & \hspace{10mm}0&39 &1.33$\pm$0.09&7.1\\
$[2.4,2.6]$ & \hspace{5mm}48&28 & \hspace{10mm}14&30 &1.35$\pm$0.09&6.4\\
$[2.6,2.85]$ & \hspace{5mm}21&20 & \hspace{10mm}35&23 &1.26$\pm$0.07&5.9\\
\hline
$[2.85,3.15]$ & \hspace{5mm}72&19 & \hspace{10mm}12&18 &1.28$\pm$0.07&5.5\\
$[3.15,3.3]$ & \hspace{5mm}19&11 & \hspace{10mm}0&3 &1.24$\pm$0.08&6.7\\
$[3.3,4]$ & \hspace{5mm}0&7 & \hspace{10mm}0&23 &1.24$\pm$0.06&4.7\\
$>$ 4 & \hspace{5mm}23&21 & \hspace{10mm}57&23 &0.94$\pm$0.05&4.9\\
\hline\hline
\end{tabular}
\end{center}
    \label{Tab:pppi_mppbins}
\end{table}

The yields and total efficiency for \pphplus in $m_{p\antiproton }$ bins are shown in Tables \ref{Tab:ppK_mppbins} and \ref{Tab:pppi_mppbins}. The charmonium contributions originate from the decays $\Bp\to \jpsi(\to p\antiproton ) \Kp$, $\Bp\to \etac(\to p\antiproton ) \Kp$ and $\Bp\to \psitwos(\to p\antiproton ) \Kp$ for the \ppkplus mode, and $\Bp\to \jpsi(\to p\antiproton ) \pip$ for the \pppiplus mode. Before deriving the distributions, the charmonium contributions are unfolded by performing two dimensional extended unbinned maximum likelihood fits to the $p\antiproton h^+$ and $p\antiproton $ invariant masses. The \jpsi and \psitwos resonances are modelled by Gaussian functions and the \etac resonance is modelled by a convolution of Breit-Wigner and Gaussian functions. The non-resonant $p\antiproton $ component and the combinatorial background are modelled by polynomial shapes. Table \ref{Tab:pph_charmonia_yields} shows the yields of contributing charmonium modes. The results are consistent with those reported in Ref.~\cite{LHCb-PAPER-2012-047}.

\begin{table}[tb]
\caption{ Yields, efficiencies and relative systematic uncertainties of the charmonium modes from the combined $(m_{p\antiproton h^+},m_{p\antiproton })$ fits for the regions $m_{p\antiproton }\in[2.85,3.15]~\gevcc$ (for both \ppkplus and \pppiplus) and $[3.60,3.75]~\gevcc$ (for \ppkplus).}
  \begin{center}
\begin{tabular}{l r@{$\pm$}l c c}\hline\hline
\centering
 Mode & \multicolumn{2}{c}{Yield} & \multicolumn{1}{c}{Efficiency (\%)} & \multicolumn{1}{c}{Syst. (\%)}\\ 
\hline 
$\Bp\to \jpsi(\to p\antiproton ) \Kp$ & 1413&40 & 1.624$\pm$0.005 & 1.8\\
$\Bp\to \etac(\to p\antiproton ) \Kp$  & 722&36 & 1.660$\pm$0.005 & 2.0\\
$\Bp\to \psitwos(\to p\antiproton ) \Kp$ & 132&16 & 1.475$\pm$0.011 & 1.5\\
\hline
$\Bp\to \jpsi(\to p\antiproton ) \pip$& 59&11 &1.328$\pm$0.011 & 4.2\\
\hline\hline
\end{tabular}
\end{center}
    \label{Tab:pph_charmonia_yields}
\end{table}

\begin{figure}[tb]%bt\centering
\begin{center}
\includegraphics[width=0.45\textwidth]{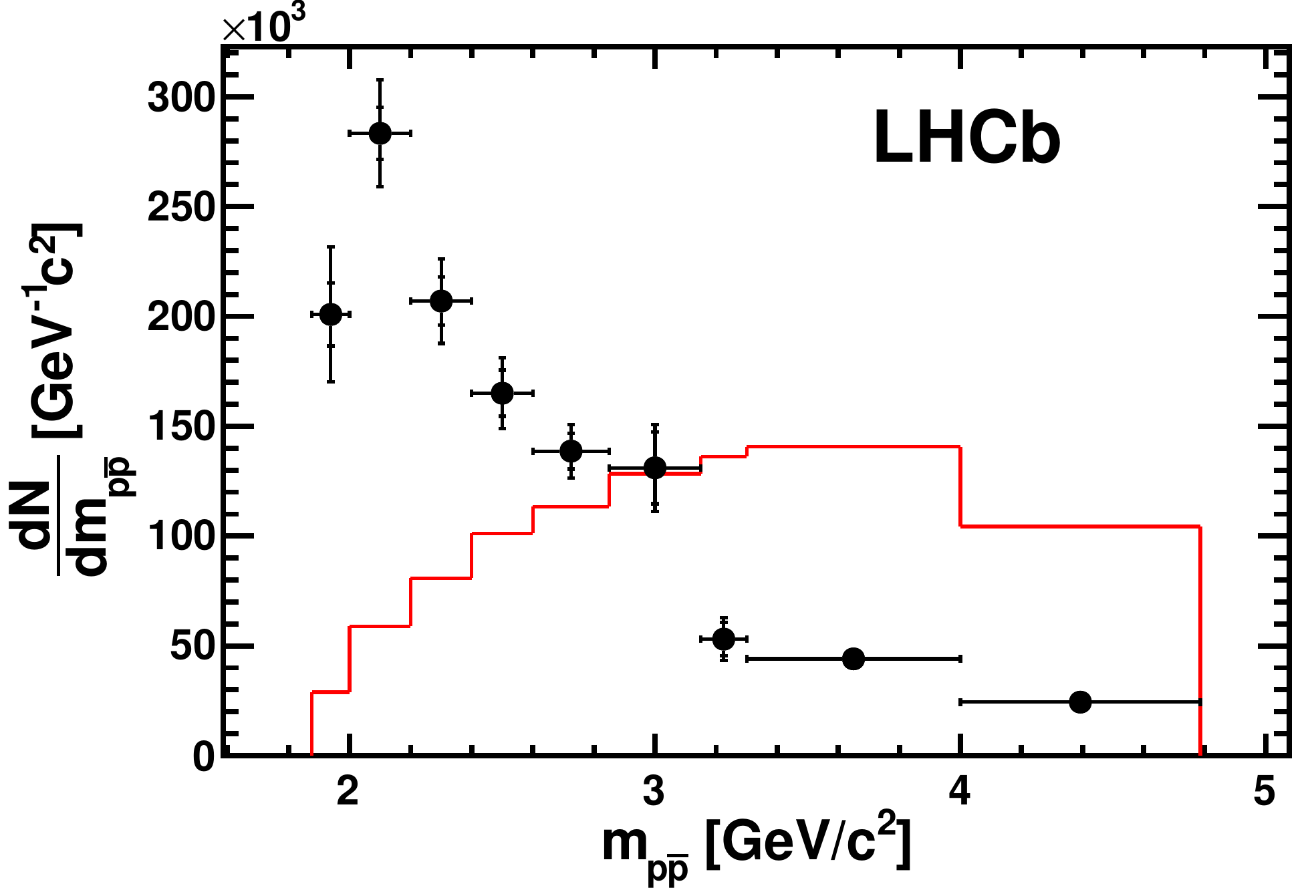}
\includegraphics[width=0.45\textwidth]{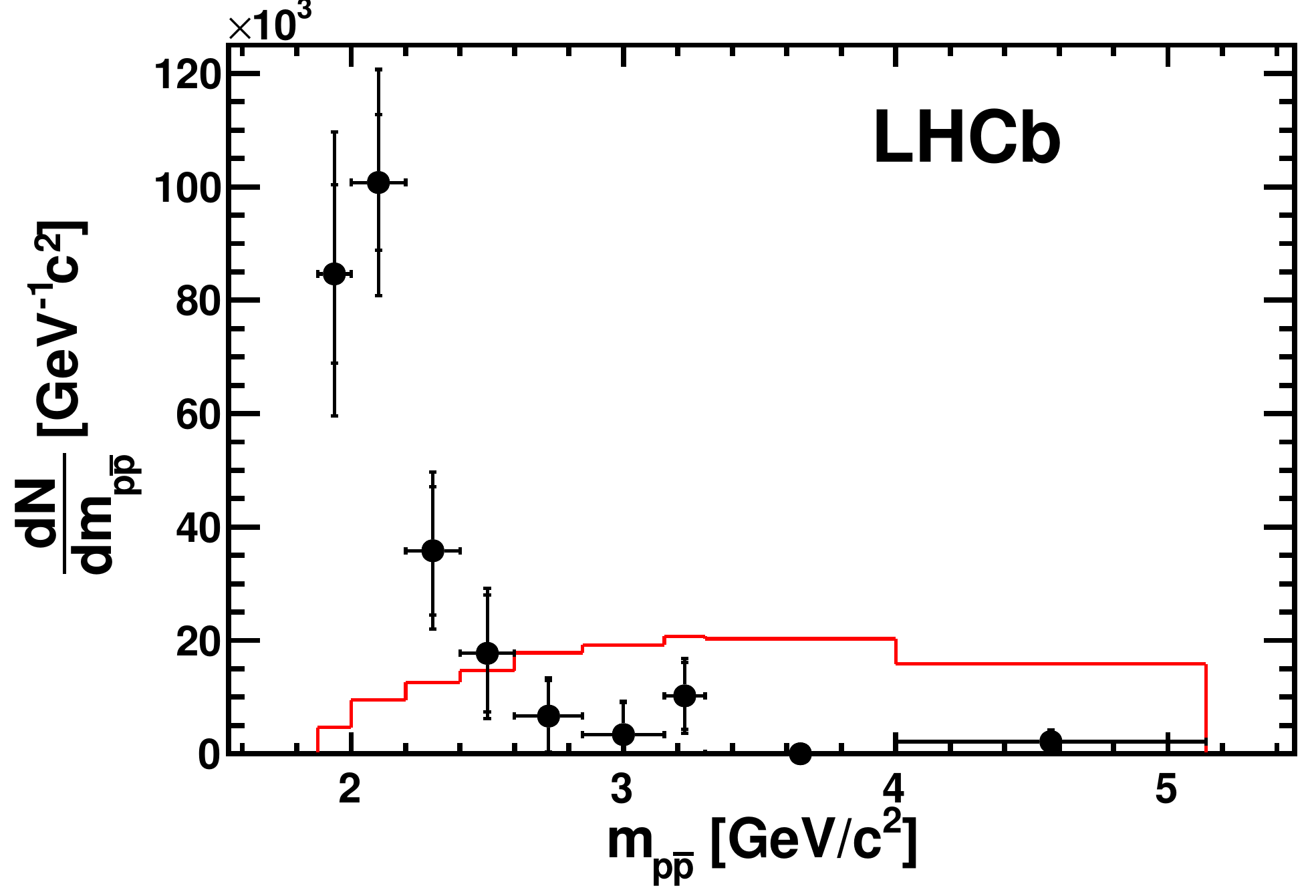}
\end{center}
\caption{Efficiency-corrected differential yield as a function of $m_{p\antiproton }$ for (left) \ppkplus and (right) \pppiplus. The data points are shown with their statistical and total uncertainties. For comparison, the solid lines represent the expectations for a uniform phase space production, normalized to the efficiency-corrected area.}
\label{Fig:pph_mppdiffspectra}
\end{figure}

After unfolding, the efficiency-corrected differential distributions are shown in Fig.~\ref{Fig:pph_mppdiffspectra}. An enhancement is observed at low $p\antiproton $ mass both for \ppkplus and \pppiplus, with a more sharply peaked distribution for \pppiplus. This accumulation of events at low $m_{p\antiproton }$ is a well known feature that has also been observed in different contexts such as $\Upsilon(1S)\to\gamma p\antiproton $ \cite{CLEO_pp}, $\jpsi\to\gamma p\antiproton $ \cite{BES_pp} and $B^0\to D^{(*)0}p\antiproton $ \cite{BaBar_dpp} decays. It appears to be caused by proton-antiproton rescattering and is modulated by the particular kinematics of the decay from which the $p\antiproton $ pair originates \cite{Haidenbauer}.

\subsection{Invariant mass squared of the \boldmath$Kp$ system}
\begin{figure}[tb]%bt\centering
\begin{center}
\includegraphics[width=0.5\textwidth]{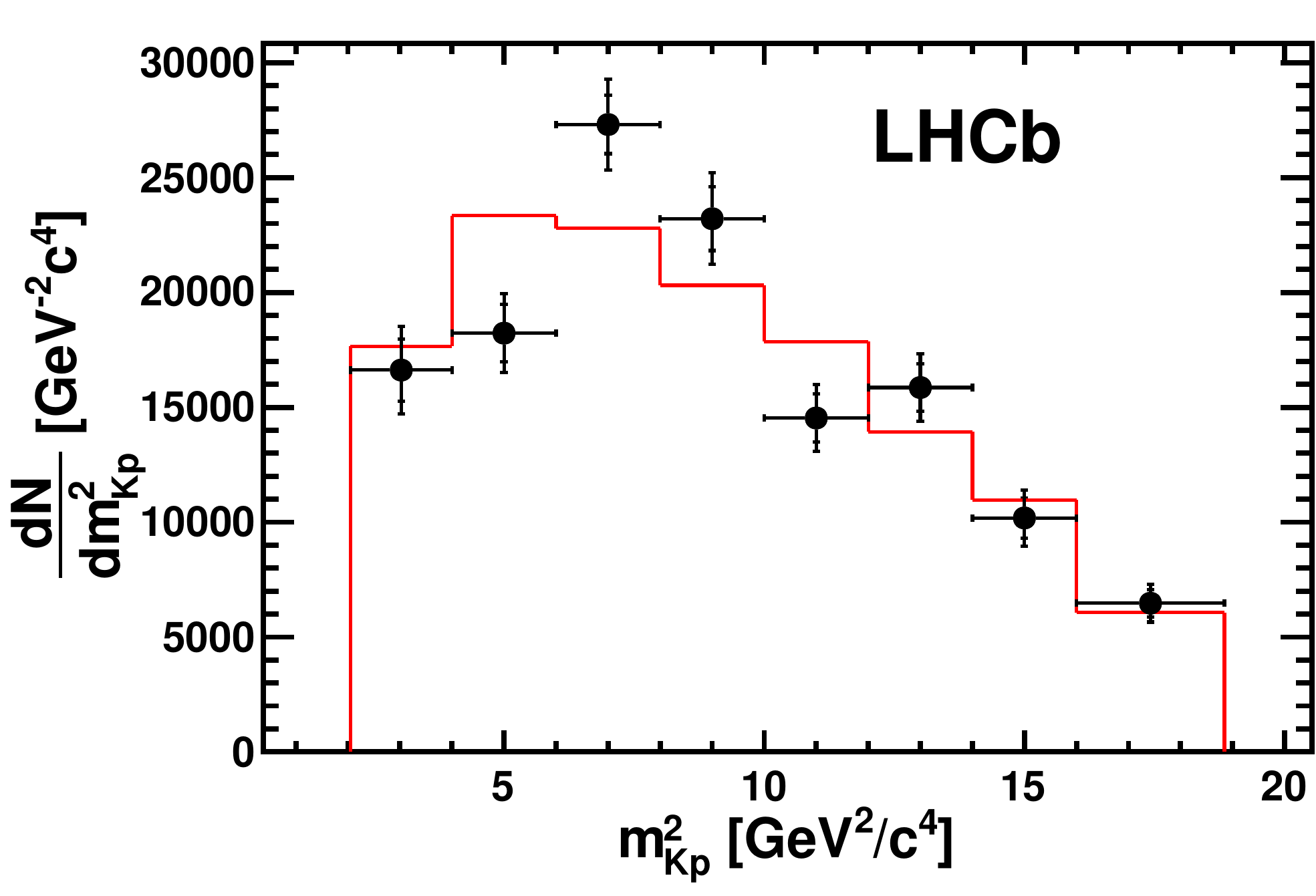}
\end{center}
\caption{Efficiency-corrected differential yield as a function of $m_{Kp}^2$ for \ppkplus. The data points are shown with their statistical and total uncertainties. The solid line represents the expectation for a uniform phase space production, normalized to the efficiency-corrected area, for comparison.}
\label{Fig:ppK_mkp2diffspectra}
\end{figure}

%where $Kp$ denotes the neutral combinations $pK^-$ or $\antiproton K^+$
The \ppkplus signal yield as a function of the Dalitz-plot variable $m_{Kp}^2$ is considered, where $Kp$ denotes the neutral combinations $K^-p$ or $K^+\antiproton $. Table \ref{Tab:ppK_mkp2bins}
 shows the yields and efficiencies, after the charmonium bands have been vetoed in the ranges $m_{p\antiproton }\in[2.85,3.15]~\gevcc$ and $[3.60,3.75]~\gevcc$. The differential spectrum derived after efficiency correction is shown in Fig.~\ref{Fig:ppK_mkp2diffspectra}. Contrary to the situation for $m_{p\antiproton }$, the data distribution is in reasonable agreement with the uniform phase space distribution, with some discrepancies in the region $m_{Kp}^2\in[4,12]~(\gevccbis)^2$.

\begin{table}[tb]
\caption{ Fitted \ppkplus yields after subtracting the charmonium bands, efficiencies and relative systematic uncertainties in bins of $Kp$ invariant mass squared.}
  \begin{center}
\begin{tabular}{cccc}\hline\hline
\centering
 $m_{Kp}^{2}$ $[(\gevcc)^{2}]$ &  \ppkplus yield & Efficiency (\%) & Syst. (\%)\\ 
\hline 
$<$ 4 & 454$\pm$37& 1.40$\pm$0.02 & 3.3\\
$[4,6]$ & 522$\pm$36&1.43$\pm$0.02 & 2.5\\
$[6,8]$ & 797$\pm$37&1.45$\pm$0.01 & 2.6\\
$[8,10]$ & 702$\pm$42 &1.51$\pm$0.01 & 2.6\\
$[10,12]$ & 445$\pm$32 &1.53$\pm$0.01 & 2.8\\
$[12,14]$ & 526$\pm$34 &1.66$\pm$0.01 & 2.8 \\
$[14,16]$ & 338$\pm$29 & 1.67$\pm$0.02 & 3.4\\
$>$ 16 & 305$\pm$28 & 1.66$\pm$0.02 & 3.5\\
\hline\hline
\end{tabular}
\end{center}
    \label{Tab:ppK_mkp2bins}
\end{table}

\subsection{Helicity angle of the \boldmath$p\antiproton $ system}
\begin{figure}[t]%bt\centering
\begin{center}
\includegraphics[width=0.45\textwidth]{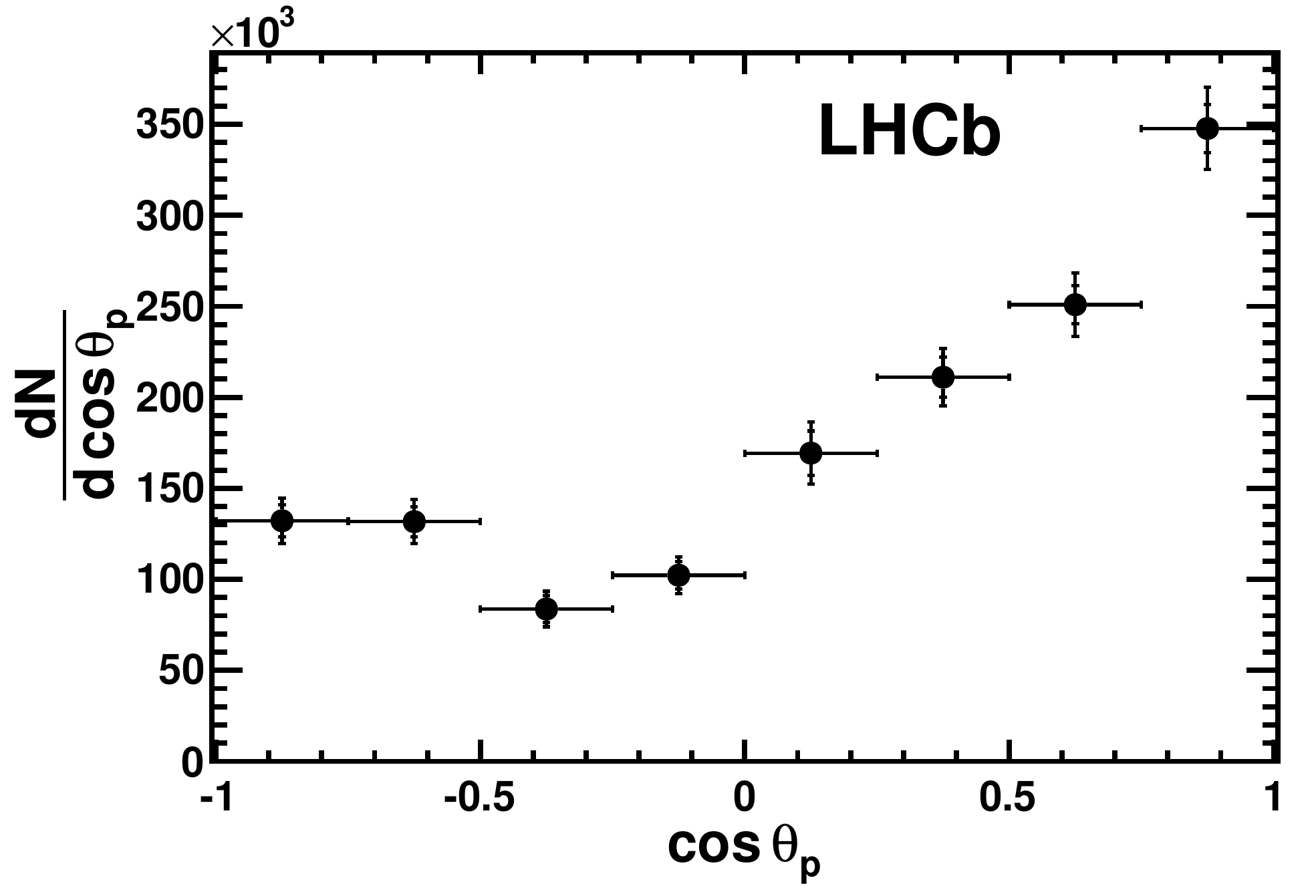}
\includegraphics[width=0.45\textwidth]{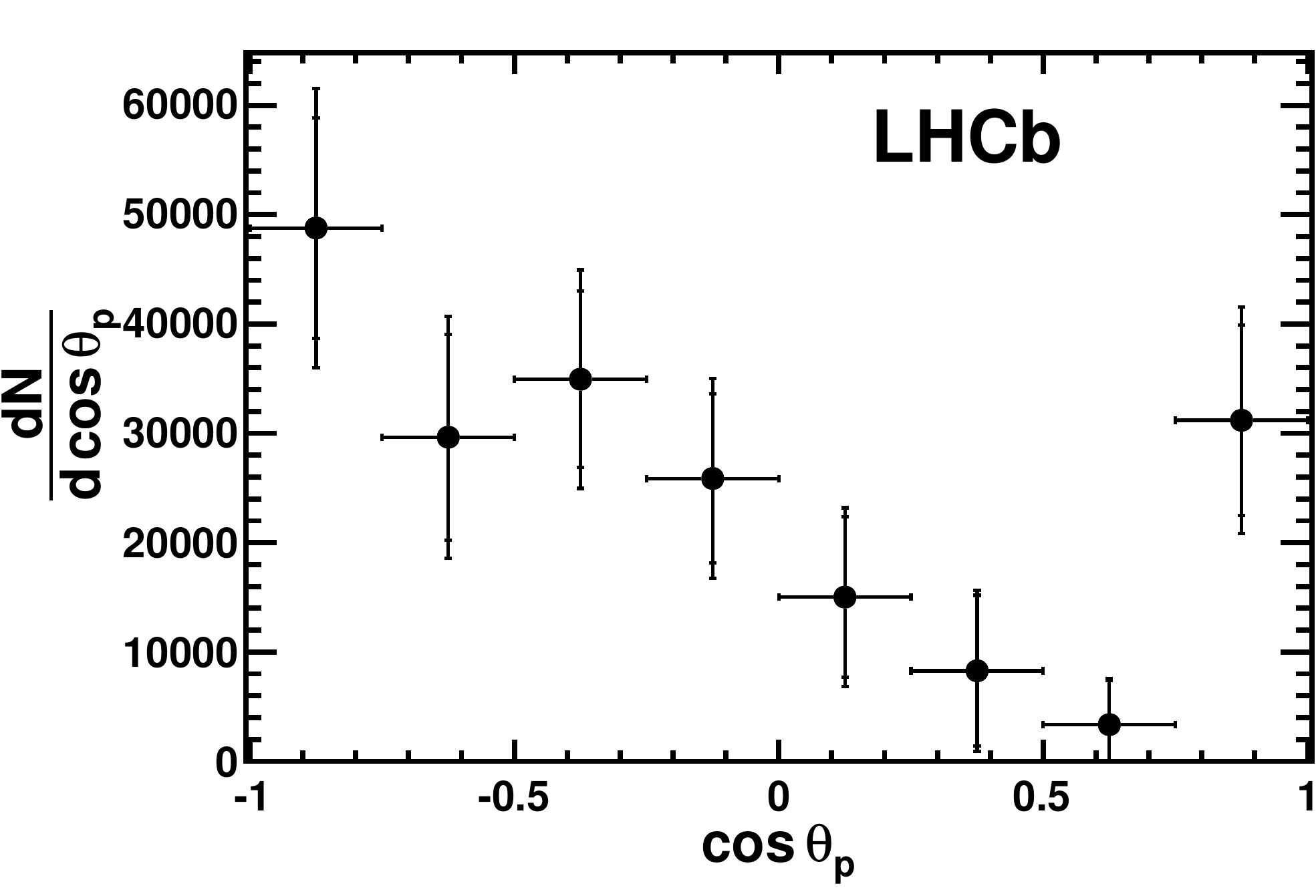}
\end{center}
\caption{Efficiency-corrected differential yields as functions of $\cos\theta_p$ for (left) \ppkplus and (right) \pppiplus modes, after subtraction of the charmonium contributions. The data points are shown with their statistical and total uncertainties.}
\label{Fig:pph_costhetadiffspectra}
\end{figure}

The \pphplus signal yields are considered as a function of $\cos\theta_p$. Tables \ref{Tab:ppK_cosbins} and \ref{Tab:pppi_cosbins} show the corresponding yields and efficiencies. The differential distributions are shown in Fig.~\ref{Fig:pph_costhetadiffspectra}.

\begin{table}[t]
\caption{ Fitted \ppkplus yields, efficiencies and relative systematic uncertainties in bins of $\cos\theta_{p}$.}
  \begin{center}
\begin{tabular}{c r@{$\pm$}l c c}\hline\hline
\centering
 $\cos\theta_{p}$ range &  \multicolumn{2}{c}{\ppkplus yield} & Efficiency (\%) & Syst. (\%)\\ 
\hline 
$[-1,-0.75]$& \hspace{7mm}508&34 &1.54$\pm$0.01&  2.7\\
$[-0.75,-0.5]$ & \hspace{7mm}497&31 &1.51$\pm$0.02 &  3.0\\
$[-0.5,-0.25]$ & \hspace{7mm}309&27 &1.48$\pm$0.01 &  2.9\\
$[-0.25,0]$ & \hspace{7mm}381&28 &1.49$\pm$0.01&  2.6\\
$[0,0.25]$ & \hspace{7mm}640&46 &1.51$\pm$0.01&  2.9\\
$[0.25,0.5]$ & \hspace{7mm}799&42 & 1.52$\pm$0.01&  2.2\\
$[0.5,0.75]$ & \hspace{7mm}976&41 &1.56$\pm$0.01&  2.8\\
$[0.75,1]$ & \hspace{7mm}1346&51 &1.55$\pm$0.01&  2.7\\
\hline\hline
\end{tabular}
\end{center}
    \label{Tab:ppK_cosbins}
\end{table}

\begin{table}[b]
\caption{Fitted \pppiplus signal yields, efficiencies and relative systematic uncertainties in bins of $\cos\theta_{p}$.}
  \begin{center}
\begin{tabular}{c r@{$\pm$}l c c}\hline\hline
\centering
 $\cos\theta_{p}$ range &  \multicolumn{2}{c}{\pppiplus yield} & Efficiency(\%) &  Syst. (\%)\\ 
\hline 
$[-1,-0.75]$& \hspace{7mm}150&31 &1.23$\pm$0.02&  5.5\\
$[-0.75,-0.5]$ & \hspace{7mm}85&27 & 1.15$\pm$0.02 &  5.5 \\
$[-0.5,-0.25]$ & \hspace{7mm}104&24 & 1.19$\pm$0.02 &  5.5\\
$[-0.25,0]$ & \hspace{7mm}77&23& 1.19$\pm$0.02&  5.5\\
$[0,0.25]$ & \hspace{7mm}43&21 & 1.14$\pm$0.02&  5.5\\
$[0.25,0.5]$ & \hspace{7mm}24&20 & 1.16$\pm$0.02&  5.5\\
$[0.5,0.75]$ & \hspace{7mm}10&12 & 1.19$\pm$0.02&  5.5\\
$[0.75,1]$ & \hspace{7mm}93&26 & 1.19$\pm$0.02&  5.2\\
\hline\hline
\end{tabular}
\end{center}
    \label{Tab:pppi_cosbins}
\end{table}

The forward-backward asymmetries are derived by comparing the yields for $\cos\theta_p>0$ and $\cos\theta_p<0$, accounting for the averaged efficiencies in each region
\begin{equation}
\AFB=\frac{\frac{N_{\mathrm{pos}}}{\epsilon_{\mathrm{pos} }}-\frac{N_{\mathrm{neg} }}{\epsilon_{\mathrm{neg} }}}{\frac{N_{\mathrm{pos}}}{\epsilon_{\mathrm{pos}}}+\frac{N_{\mathrm{neg}}}{\epsilon_{\mathrm{neg}}}}=\frac{N_{\mathrm{pos}}-fN_{\mathrm{neg}}}{N_{\mathrm{pos}}+fN_{\mathrm{neg}}},
\end{equation}
where $\epsilon_{\mathrm{pos}}=\epsilon(\cos\theta_p>0)$ and $\epsilon_{\mathrm{neg}}=\epsilon(\cos\theta_p<0)$ are the averaged efficiencies, $f=\epsilon_{\mathrm{pos}}/\epsilon_{\mathrm{neg}}$ and $N_{\mathrm{pos}}=N(\cos\theta_p>0)$, $N_{\mathrm{neg}}=N(\cos\theta_p<0)$. The values obtained are: $\AFB(p\antiproton \Kp)=0.370\pm0.018~(\mathrm{stat})\pm0.016~(\mathrm{syst})$ and $\AFB(p\antiproton \pip)=-0.392\pm0.117~(\mathrm{stat})\pm0.015~(\mathrm{syst})$.

A clear opposite angular correlation between \ppkplus and \pppiplus decays is observed; the light meson $h$ tends to align with the opposite-sign baryon for \ppk while it aligns with the same-sign baryon for the \pppi mode. A quark level analysis suggests that the meson should align with the same-sign baryon, since the opposite-sign baryon has larger momentum, being formed by products from the decaying \b quark \cite{HYCheng}. This is in agreement with the angular spectrum of \pppiplus but not for \ppkplus decays. 

\subsection{Dalitz plot}

\begin{figure}[tb]%bt\centering
\begin{center}
\includegraphics[width=0.45\textwidth]{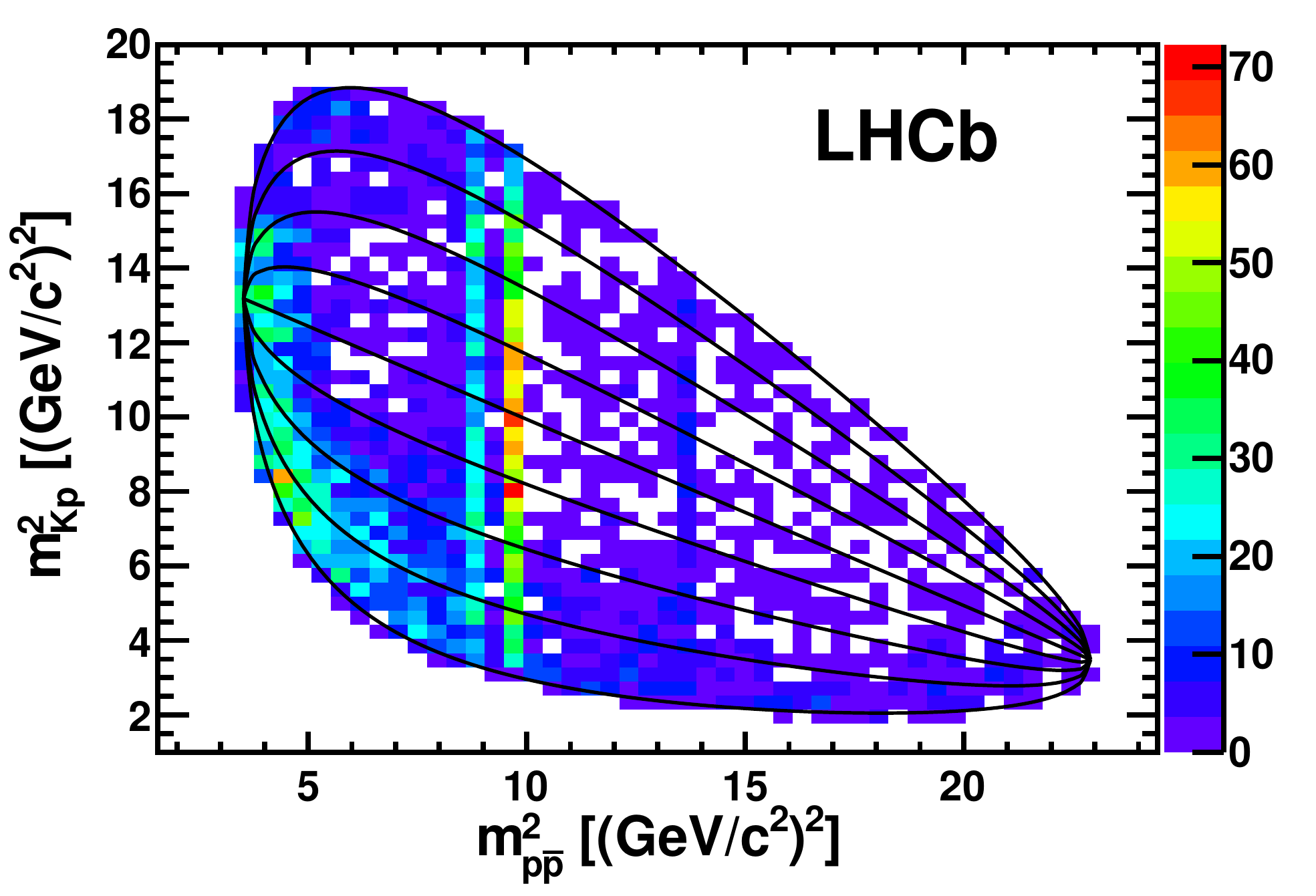}
\includegraphics[width=0.45\textwidth]{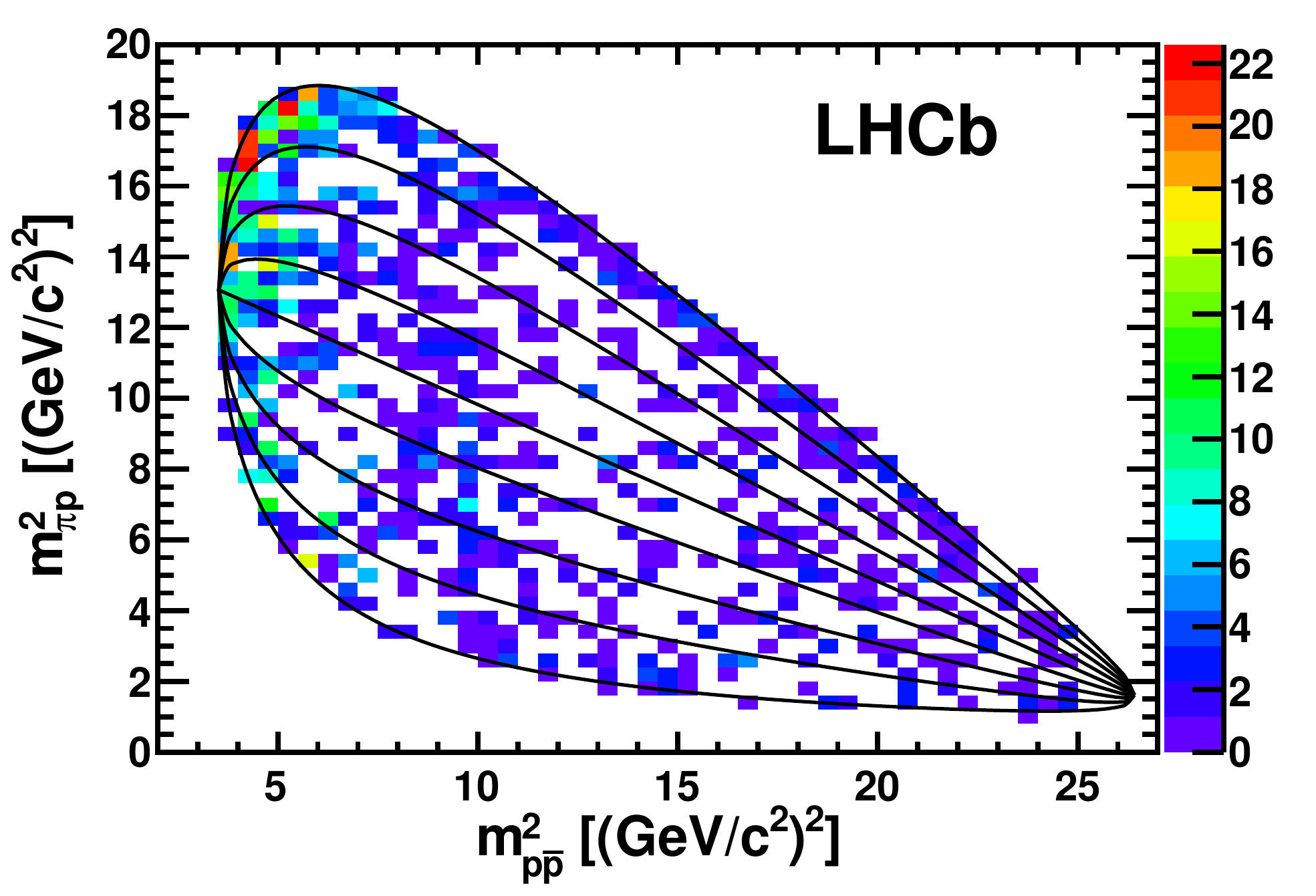}
\end{center}
\caption{Signal weighted Dalitz-plot distributions for (left) \ppkplus and (right) \pppiplus. Also shown are the iso-$\cos\theta_p$ lines corresponding to the $\cos\theta_p$ bin boundaries; $\cos\theta_p=-1$ (+1) is the uppermost (lowermost) line. The distributions are not corrected for efficiency.}
\label{Fig:pph_dalitz_splots}
\end{figure}

From the fits to the $B$-candidate invariant mass, shown in Fig.~\ref{Fig:pph_globalfits}, signal weights are calculated with the {\it sPlot} technique \cite{sPlots_NIM} and are used to produce the signal Dalitz-plot distributions shown in Fig.~\ref{Fig:pph_dalitz_splots}. To ease the comparison, the $\cos\theta_p$ curves corresponding to the boundaries of the eight bins used to make the angular distributions in Fig.~\ref{Fig:pph_costhetadiffspectra} are superimposed.

With the exception of the charmonium bands (\etac, \jpsi, \psitwos for \ppkplus, and \jpsi for \pppiplus), the structure of the low $p\antiproton $ mass enhancement is very different between \ppkplus and \pppiplus.
The \ppkplus events are distributed in the middle and lower $m_{Kp}^2$ half, exhibiting a possible $p\antiproton $ band structure near $4~\mathrm{GeV}^2/c^4$. An enhancement at low $m_{Kp}$ is also observed and is caused to a large extent by a $\Lz(1520)$ signal, as will be shown in the next section.
The \pppiplus events are mainly clustered in the upper $m_{\pi p}^2$ half, with also a few events on the doubly-charged top diagonal $(p\pi)^{++}$ (near the $\cos\theta_p=-1$ boundary).
These distributions of events are consistent with the angular distributions and asymmetries reported earlier.

\section{Measurement of \boldmath\acp for \ppkplus decays}
The raw charge asymmetry is obtained by performing a simultaneous extended unbinned maximum likelihood fit to the $B^-$ and $B^+$ samples. The \Bpm yields are defined as a function of the total yield $N$ and the raw asymmetry, \acpraw, by $N^\mp=N(1\pm\acpraw)/2$.

The \CP asymmetry is then derived after correcting for the \Bpm production asymmetry $\aprod(\Bpm)$ and the kaon detection asymmetry $\adet(\Kpm)$
\begin{equation}
\acp = \acpraw - \aprod(\Bpm) - \adet(\Kpm).
\end{equation}
The correction $A_\Delta=\aprod(\Bpm) + \adet(\Kpm)$ is measured from data with the decay $\Bpm\to \jpsi(\to p\antiproton ) \Kpm$ which is part of the data sample
\begin{equation}
A_\Delta = \acpraw( \jpsi(\to p\antiproton ) \Kpm)-\acp(\jpsi\Kpm),
\end{equation}
where $\acp(\jpsi\Kpm)=(1\pm7)\times10^{-3}$~\cite{PDG}.

Another correction has been applied to account for the proton antiproton asymmetry, which exactly cancels for $\jpsi(\to p\antiproton ) \Kpm$ but not necessarily in the full phase space of $p\antiproton \Kpm$ events. This effect has been estimated in simulation studying the difference in the interactions of protons and antiprotons with the detector material between $\jpsi(\to p\antiproton ) \Kpm$ and $p\antiproton \Kpm$ events generated uniformly over phase space. We obtained a $m_{Kp}^2$-dependent bias, up to 3\% for the highest bin, for \acpraw.

\begin{figure}[tb]%bt\centering
\begin{center}
\includegraphics[width=0.45\textwidth]{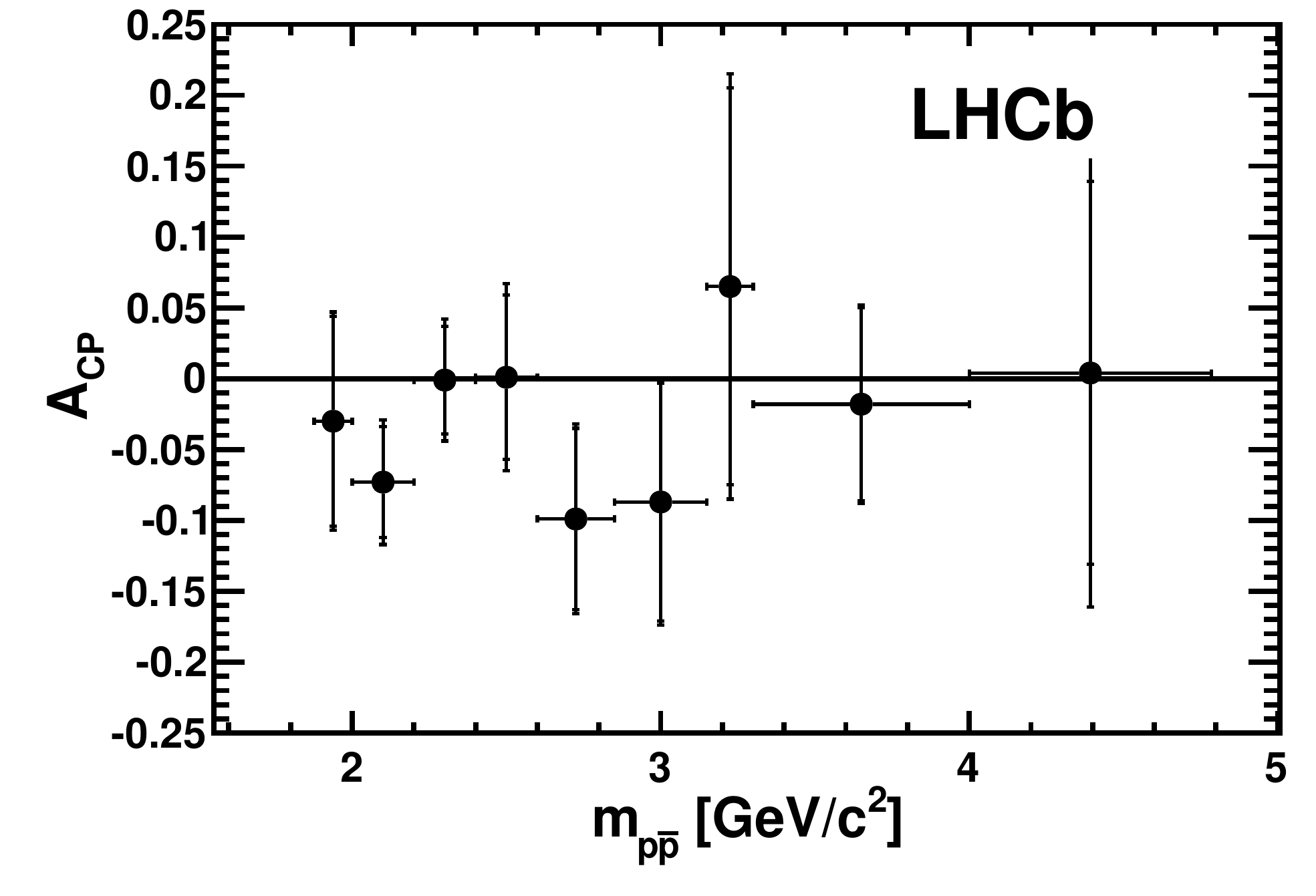}%[scale=.45]
\includegraphics[width=0.45\textwidth]{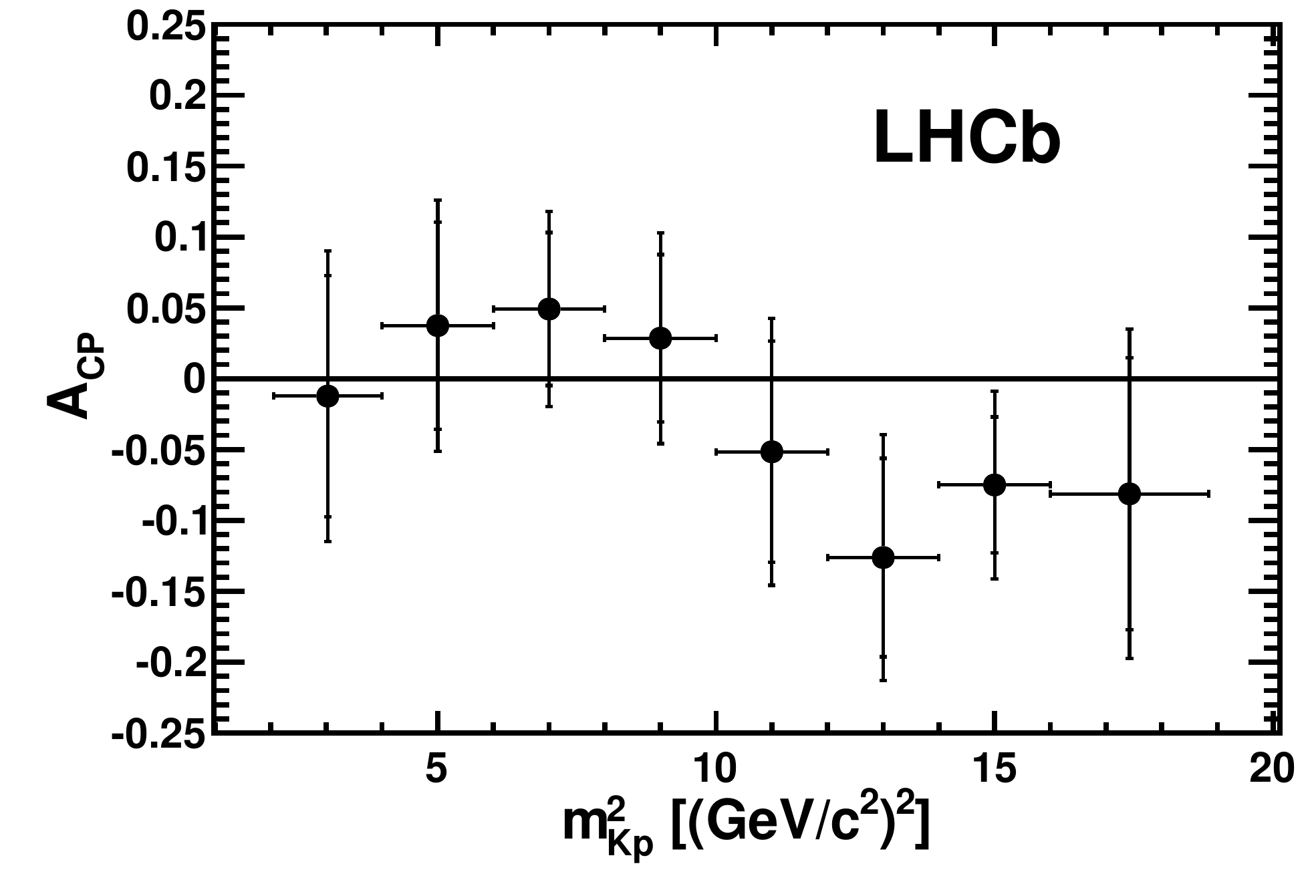}%[scale=.45]
\end{center}
\caption{Distribution of \acp for the Dalitz-plot projections on $m_{p\antiproton }$ and $m_{Kp}^2$ for \ppk events. In the $m_{p\antiproton }$ projection (left), the bin $[2.85,3.15]~\gevcc$ contains only the value of the charmless $p\antiproton \Kpm$ after subtraction of the \etac-\jpsi contribution. The $m_{Kp}^2$ projection (right) has been obtained after removing the charmonia bands.}
\label{Fig:acp_ppk_distrib}
\end{figure}

To measure \acpraw for charmonium modes, and in particular $\jpsi(\to p\antiproton ) \Kpm$, a two dimensional $(m_B,m_{p\antiproton })$ simultaneous fit to the $B^+$ and $B^-$ samples is performed. The systematic uncertainties are estimated by varying the fit functions and splitting the data sample according to trigger requirements or magnet polarities, and recombining the results from the sub-samples. The procedure is applied to obtain a global value of \acp as well as the variation of the asymmetry as a function of the Dalitz-plot variables.
The results are: $\acp=-0.022\pm0.031~(\mathrm{stat})\pm0.007~(\mathrm{syst})$ for the full $p\antiproton \Kpm$ spectrum, and $\acp=-0.047\pm0.036~(\mathrm{stat})\pm0.007~(\mathrm{syst})$ for the region $m_{p\antiproton }<2.85~\gevcc$. Figure \ref{Fig:acp_ppk_distrib} shows the variation of \acp as a function of the Dalitz-plot variables.

For the charmonium resonances, the values are: $\acp(\etac\Kpm)=0.046\pm0.057~(\mathrm{stat})\pm0.007~(\mathrm{syst})$ and $\acp(\psitwos\Kpm)=-0.002\pm0.123~(\mathrm{stat})\pm0.012~(\mathrm{syst})$. All results indicate no significant \CP asymmetries.

\section{Observation of the \boldmath$\lambdaFifteenTwentypplus$ decay}
\begin{figure}[tb]%bt\centering
\begin{center}
\includegraphics[width=0.5\textwidth]{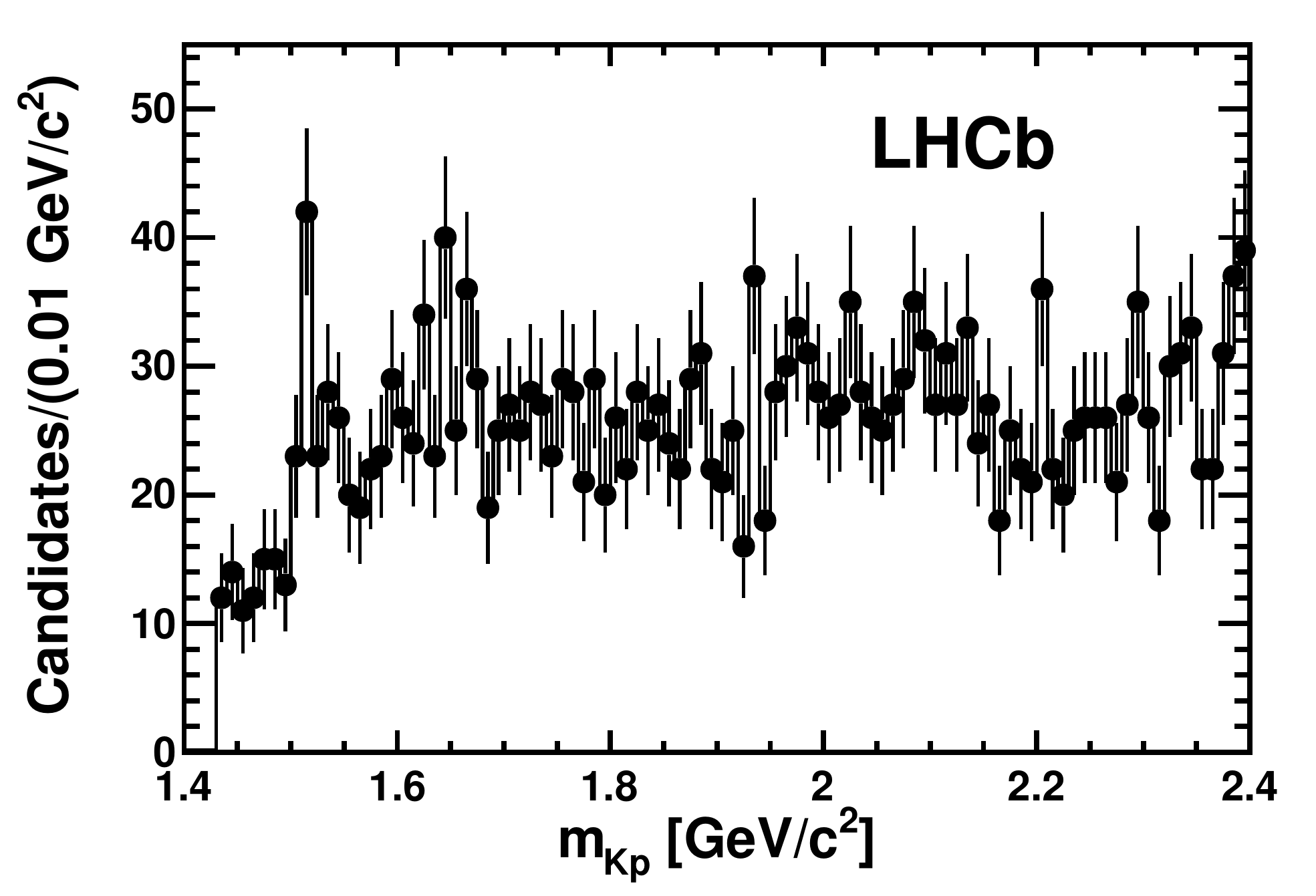}
\end{center}
\caption{Invariant mass $m_{Kp}$ for the \ppkplus candidates near threshold.}
\label{Fig:ppK_mkprawspec}
\end{figure}

In the $p\antiproton \Kp$ spectrum, near the threshold of the neutral $Kp$ combination, a peak in invariant mass at 1.52\gevcc is observed, as shown in Fig.~\ref{Fig:ppK_mkprawspec}, corresponding to the $\uquarkbar\dquarkbar\squarkbar$ resonance $\Lbar(1520)$. The possible presence of higher $\Lbar$ and $\Sigmaresbar$ resonances may explain the enhancement in the range of $[1.6,1.7]~\gevcc$.

\begin{figure}[tb]%bt\centering
\begin{center}
\includegraphics[width=0.5\textwidth]{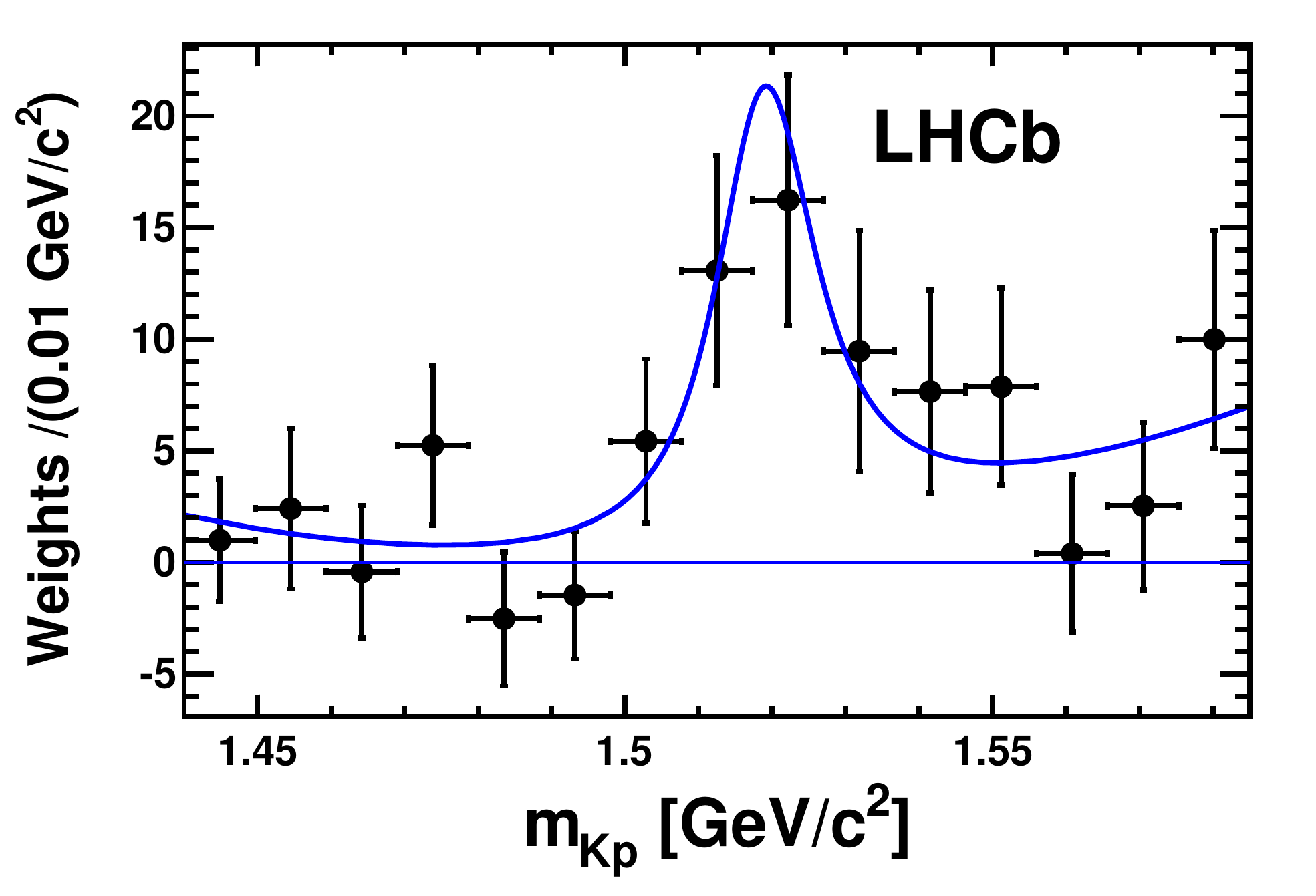}
\end{center}
\caption{Fit to the $B$ signal weighted $m_{Kp}$ distribution.}
\label{Fig:ppK_lambda1520_sw}
\end{figure}

To identify the $\Lbar(1520)$ signal, the $B^+$ signal is analyzed in the region $m_{Kp}\in[1.44,1.585]~\gevcc$. Figure \ref{Fig:ppK_lambda1520_sw} shows the $B$ signal weighted $Kp$ invariant mass, and the expected $\Lbar(1520)$ shape obtained from a model based on an asymmetric Breit-Wigner function derived from an \evtgen~\cite{Lange:2001uf} simulation of the decay \lambdaFifteenTwentypplus, convolved with a Gaussian resolution function, and a second-order polynomial function representing the tail of the non-$\Lbar(1520)$ \ppkplus decays. 

\begin{figure}[tb]%bt\centering
\begin{center}
\includegraphics[width=0.45\textwidth]{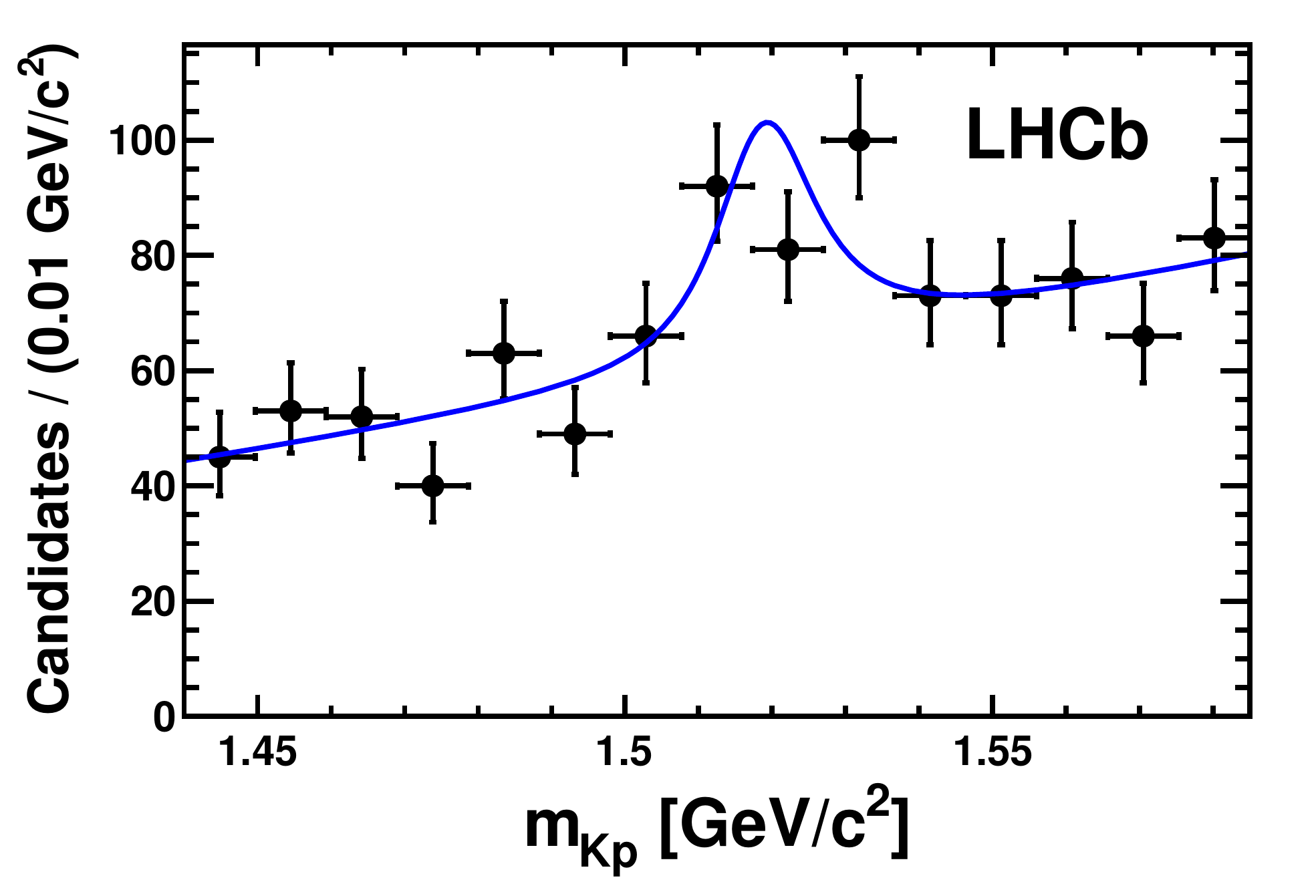}
\includegraphics[width=0.45\textwidth]{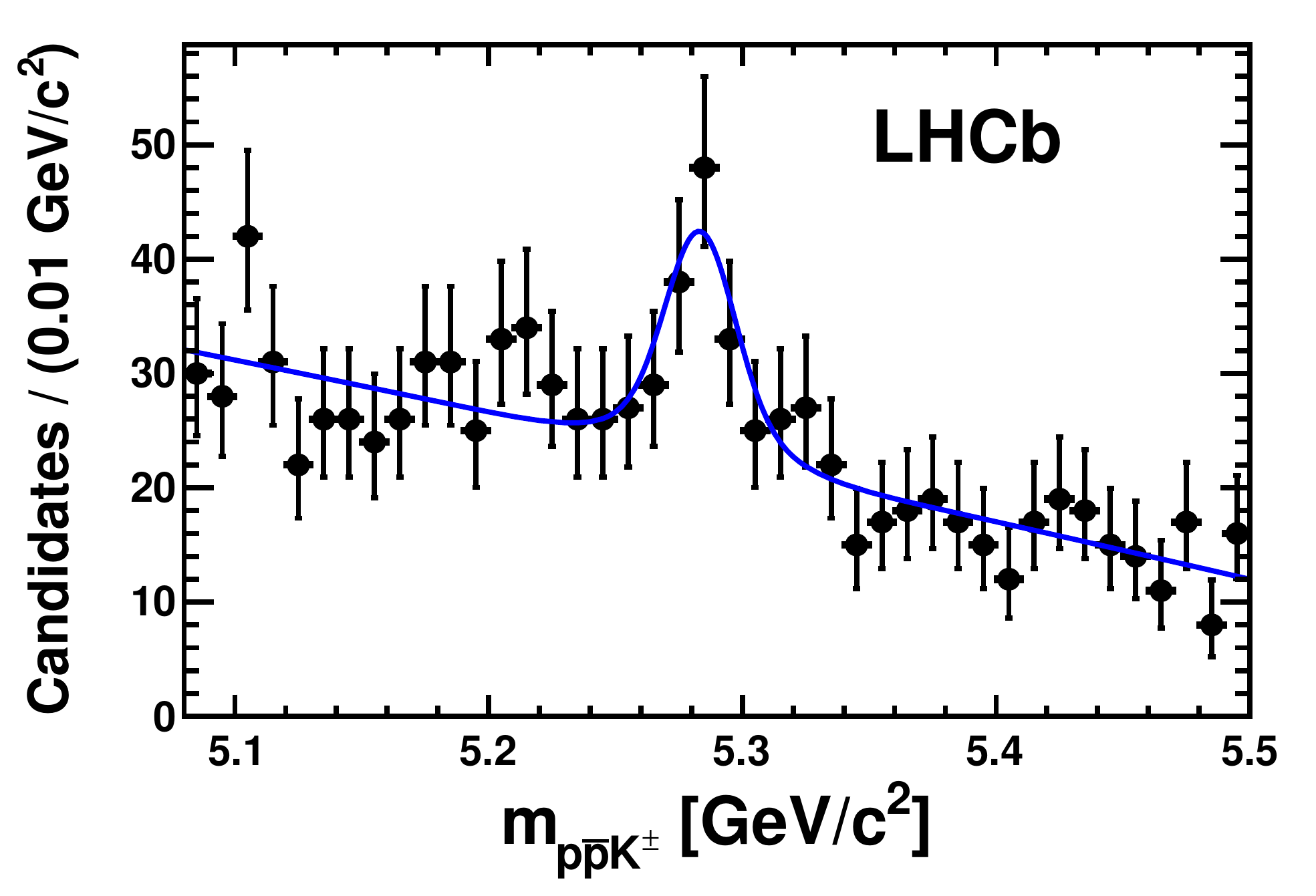}
\end{center}
\caption{Projections of (left) $m_{Kp}$ and (right) $m_{p\antiproton \Kp}$ of the two dimensional fit used to obtain the \lambdaFifteenTwentypplus signal yield.}
\label{Fig:ppK_lambda1520_sig}
\end{figure}

%\caption{(left) Projection of $m_{p\antiproton \Kp}$ of the two dimensional fit used to obtain the \lambdaFifteenTwentypplus signal yield and (right) fit to the $B$ signal weighted $m_{Kp}$ distribution, with the $\Lambda(1520)$ shape and the upper side-band tail (see text for details).}
These shapes are then used in a two dimensional $(m_{p\antiproton \Kp},m_{Kp})$ extended unbinned maximum likelihood fit to obtain the \lambdaFifteenTwentypplus yield. The fit results in $N(\lambdaFifteenTwentypplus)=47^{+12}_{-11}$ with a statistical significance of $5.3$ standard deviations, obtained by comparing the likelihood at its maximum for the nominal fit and for the background-only hypothesis. Figure \ref{Fig:ppK_lambda1520_sig} shows the projections of the fit for the $Kp$ and $p\antiproton \Kp$ invariant masses.

To test the robustness of the observation, different representations of the $Kp$ background have been used, combining first or second order polynomials and a contribution modelled by a Breit-Wigner function, for which the mean ($\mu$) and width ($\Gamma$) are allowed to vary within the known values of the $\Lz(1600)$ baryon ($\mu\in[1.56,1.7]~\gevcc$, $\Gamma\in[0.05,0.25]~\gevcc$). Fits in a wider $m_{Kp}$ range were also considered. In all cases the yield was stable with a statistical significance similar to the nominal fit case. 

The branching fraction for the decay \lambdaFifteenTwentypplus is derived from the ratio

\begin{equation}
\frac{ \mathcal{B}(\Bp\to\Lbar(1520)(\to K^+\antiproton )p) }{\mathcal{B}(\Bp \to \jpsi(\to p\antiproton )\Kp)}= \frac{N_{\Lz(1520)\to Kp}}{N_{\jpsi\to p\antiproton }}\times\frac{\epsilon_{\jpsi\to p\antiproton }^{\mathrm{gen}}}{\epsilon_{\Lz(1520)\to Kp}^{\mathrm{gen}} }\times\frac{\epsilon_{\jpsi\to p\antiproton }^{\mathrm{sel}}}{\epsilon_{\Lz(1520)\to Kp}^{\mathrm{sel}} }, 
\end{equation}
where $N_i$ is the yield of the decay chain $i$, $\epsilon^{\mathrm{gen}}$ denotes the efficiency after geometrical acceptance and simulation requirements.  The global selection efficiency $\epsilon^{\mathrm{sel}}$ includes the reconstruction, the trigger, the offline selection, and the particle identification requirements. 
The ratio of branching fractions obtained is
\begin{equation}
\nonumber \frac{ \mathcal{B}(\Bp\to\Lbar(1520)(\to K^+\antiproton )p) }{\mathcal{B}(\Bp \to \jpsi(\to p\antiproton )\Kp)}= 0.041^{+0.011}_{-0.010}~(\mathrm{stat})\pm 0.001~(\mathrm{syst}).
\end{equation}
The systematic uncertainties include effects of the $Kp$ background model, the particle identification, the limited simulation sample size, the uncertainties on the relative trigger efficiencies, and are summarized in Table \ref{lambda1520_syst}. Convolving the systematic uncertainty with the statistical likelihood profile, the global significance is 5.1 standard deviations.

\begin{table}[tb]
\caption{Systematic uncertainties for the $\mathcal{B}(\Bp\to\Lbar(1520)(\to K^+\antiproton )p)/\mathcal{B}(\Bp \to \jpsi(\to p\antiproton )\Kp)$  branching fraction ratio. The total uncertainty is the sum in quadrature of the individual sources.}
  \begin{center}
\begin{tabular}{lc}\hline\hline
\centering
 Source & Uncertainty (\%)\\ 
\hline 
$Kp$ background & 2.1\\
PID & 1.7\\
Simulation sample size & 0.5\\
Trigger & 1.0\\
\hline
Total & 2.9\\
\hline\hline
\end{tabular}
\end{center}
    \label{lambda1520_syst}
\end{table}

Using $\mathcal{B}(\Bp \to \jpsi\Kp)= (1.016\pm0.033)\times 10^{-3}$, $\mathcal{B}(\jpsi\to p\antiproton )=(2.17\pm0.07)\times 10^{-3}$ \cite{PDG}, and $\mathcal{B}(\Lambda(1520)\to K^-p)=0.234\pm0.016$ \cite{Saphir}, the branching fraction is
\begin{center}
$\mathcal{B}(\lambdaFifteenTwentypplus)=(3.9^{+1.0}_{-0.9}~(\mathrm{stat})\pm0.1~(\mathrm{syst})\pm0.3~(\mathrm{BF}))\times 10^{-7}$.
\end{center}
The last error corresponds to the uncertainty on the secondary branching fractions. This result is in agreement with the upper limit set in Ref.~\cite{BaBar_pph1}, $\mathcal{B}(\lambdaFifteenTwentypplus)<1.5\times10^{-6}$.\\
Considering the separate \Bpm signals in the range $m_{Kp}\in[1.44,1.585]~\gevcc$, the yields are $N(B^-)=50\pm12$ and $N(B^+)=27\pm11$.

\section{Summary}
Based on a data sample, corresponding to an integrated luminosity of 1.0\invfb, collected in 2011 by the \lhcb experiment, an analysis of the three body \pphplus decays ($h=K$ or $\pi$) has been performed. The dynamics of the decays has been probed using differential spectra of Dalitz-plot variables and signal-weighted Dalitz plots. The charmless \ppkplus decay populates mainly the low $m_{p\antiproton }^2$ and lower $m_{K^+\antiproton }^2$-half regions whereas the \pppiplus decay has a similar enhancement at low $m_{p\antiproton }^2$ but with an upper $m_{\pi^+\antiproton }^2$-half occupancy. From the occupation pattern of the Dalitz plots, it is likely that the \ppkplus decay is primarily driven by $p\antiproton $ rescattering with a secondary contribution from neutral $Kp$ rescattering while the \pppiplus decay is also dominated by $p\antiproton $ rescattering but with a secondary contribution from doubly-charged $(p\pi)^{++}$ rescattering, along the lines of the rescattering amplitude analysis performed in Ref.~\cite{Laporta}.
This difference of behaviour is reflected in the values of the forward-backward asymmetry of the light meson in the $p\antiproton $ rest frame
\begin{center}
$\AFB(p\antiproton \Kp)=\phantom{-}0.370\pm0.018~(\mathrm{stat})\pm0.016~(\mathrm{syst})$,\\
\hspace{1mm}$\AFB(p\antiproton \pip)=-0.392\pm0.117~(\mathrm{stat})\pm0.015~(\mathrm{syst})$.
\end{center}
\CP asymmetries for the \ppkplus decay have been measured and no significant deviation from zero observed: $\acp=-0.047\pm0.036~(\mathrm{stat})\pm0.007~(\mathrm{syst})$ for the charmless region $m_{p\antiproton }<2.85~\gevcc$,  $\acp(\etac\Kpm)=0.046\pm0.057~(\mathrm{stat})\pm0.007~(\mathrm{syst})$ and $\acp(\psitwos\Kpm)=-0.002\pm0.123~(\mathrm{stat})\pm0.012~(\mathrm{syst})$. These measurements are consistent with the current known values, $\acp(\ppk,m_{p\antiproton }<2.85~\gevcc)=-0.16\pm0.07$ \cite{PDG}, $\acp(\etac\Kpm)=-0.16\pm0.08~(\mathrm{stat})\pm0.02~(\mathrm{syst})$ \cite{Belle_pph}, and $\acp(\psitwos\Kpm)=-0.025\pm0.024$ \cite{PDG}.
The absence of any significant charge asymmetry, contrary to the situation for \hhhplus decays \cite{LHCb-CONF-2012-028,LHCb-PAPER-2013-027}, may be due to different long range behaviour. Final state interactions in the \pphplus case do not change the nature of the particles, such as $p\antiproton \to p\antiproton $ or $ph\to ph$, while \hhhplus modes can be affected by $\pi^+\pi^- \leftrightarrow K^+K^-$ scattering.

Finally, the observation of the decay \lambdaFifteenTwentypplus is reported, with the branching fraction 
\begin{center}
$\mathcal{B}(\lambdaFifteenTwentypplus)=(3.9^{+1.0}_{-0.9}~(\mathrm{stat})\pm0.1~(\mathrm{syst})\pm0.3~(\mathrm{BF}))\times 10^{-7}$,
\end{center}
in agreement with the current existing upper limit \cite{BaBar_pph1}.

\section*{Acknowledgements}

\noindent We express our gratitude to our colleagues in the CERN
accelerator departments for the excellent performance of the LHC. We
thank the technical and administrative staff at the LHCb
institutes. We acknowledge support from CERN and from the national
agencies: CAPES, CNPq, FAPERJ and FINEP (Brazil); NSFC (China);
CNRS/IN2P3 and Region Auvergne (France); BMBF, DFG, HGF and MPG
(Germany); SFI (Ireland); INFN (Italy); FOM and NWO (The Netherlands);
SCSR (Poland); MEN/IFA (Romania); MinES, Rosatom, RFBR and NRC
``Kurchatov Institute'' (Russia); MinECo, XuntaGal and GENCAT (Spain);
SNSF and SER (Switzerland); NAS Ukraine (Ukraine); STFC (United
Kingdom); NSF (USA). We also acknowledge the support received from the
ERC under FP7. The Tier1 computing centres are supported by IN2P3
(France), KIT and BMBF (Germany), INFN (Italy), NWO and SURF (The
Netherlands), PIC (Spain), GridPP (United Kingdom). We are thankful
for the computing resources put at our disposal by Yandex LLC
(Russia), as well as to the communities behind the multiple open
source software packages that we depend on.

\addcontentsline{toc}{section}{References}
\setboolean{inbibliography}{true}
\bibliographystyle{LHCb}
\bibliography{main,LHCb-PAPER,LHCb-CONF,LHCb-DP}

\ifx\mcitethebibliography\mciteundefinedmacro
\PackageError{LHCb.bst}{mciteplus.sty has not been loaded}
{This bibstyle requires the use of the mciteplus package.}\fi
\providecommand{\href}[2]{#2}
\begin{mcitethebibliography}{10}
\mciteSetBstSublistMode{n}
\mciteSetBstMaxWidthForm{subitem}{\alph{mcitesubitemcount})}
\mciteSetBstSublistLabelBeginEnd{\mcitemaxwidthsubitemform\space}
{\relax}{\relax}

\bibitem{LHCb-CONF-2012-028}
{LHCb collaboration}, \ifthenelse{\boolean{articletitles}}{{\it {Evidence for
  \CP violation in $B \to KK\pi$ and $B \to \pi\pi\pi$ decays}}, }{}
  \href{http://cdsweb.cern.ch/search?p={LHCb-CONF-2012-028}&f=reportnumber&action_search=Search&c=LHCb+Reports&c=LHCb+Conference+Proceedings&c=LHCb+Conference+Contributions&c=LHCb+Notes&c=LHCb+Theses&c=LHCb+Papers}
  {{LHCb-CONF-2012-028}}\relax
\mciteBstWouldAddEndPuncttrue
\mciteSetBstMidEndSepPunct{\mcitedefaultmidpunct}
{\mcitedefaultendpunct}{\mcitedefaultseppunct}\relax
\EndOfBibitem
\bibitem{LHCb-PAPER-2013-027}
LHCb collaboration, R.~Aaij {\em et~al.},
  \ifthenelse{\boolean{articletitles}}{{\it {\CP violation in the phase space
  of $B^\pm \to K^\pm\pi^+\pi^-$ and $B^\pm \to K^\pm K^+K^-$}},
  }{}\href{http://arxiv.org/abs/1306.1246}{{\tt arXiv:1306.1246}}, {submitted
  to Phys. Rev. Lett.}\relax
\mciteBstWouldAddEndPunctfalse
\mciteSetBstMidEndSepPunct{\mcitedefaultmidpunct}
{}{\mcitedefaultseppunct}\relax
\EndOfBibitem
\bibitem{Riazuddin}
R.~Marshak, Riazuddin, and C.~Ryan, {\em {Theory of weak interactions in
  particle physics}}, Wiley-Interscience, New York, NY, USA, 1969\relax
\mciteBstWouldAddEndPuncttrue
\mciteSetBstMidEndSepPunct{\mcitedefaultmidpunct}
{\mcitedefaultendpunct}{\mcitedefaultseppunct}\relax
\EndOfBibitem
\bibitem{Wolfenstein}
L.~Wolfenstein, \ifthenelse{\boolean{articletitles}}{{\it {Final state
  interactions and CP violation in weak decays}},
  }{}\href{http://dx.doi.org/10.1103/PhysRevD.43.151}{Phys.\ Rev.\  {\bf D43}
  (1991) 151}\relax
\mciteBstWouldAddEndPuncttrue
\mciteSetBstMidEndSepPunct{\mcitedefaultmidpunct}
{\mcitedefaultendpunct}{\mcitedefaultseppunct}\relax
\EndOfBibitem
\bibitem{Compound}
H.~Y. Cheng, C.~K. Chua, and A.~Soni, \ifthenelse{\boolean{articletitles}}{{\it
  {Final state interactions in hadronic $B$ decays}},
  }{}\href{http://dx.doi.org/10.1103/PhysRevD.71.014030}{Phys.\ Rev.\  {\bf
  D71} (2005) 014030}, \href{http://arxiv.org/abs/hep-ph/0409317}{{\tt
  arXiv:hep-ph/0409317}}\relax
\mciteBstWouldAddEndPuncttrue
\mciteSetBstMidEndSepPunct{\mcitedefaultmidpunct}
{\mcitedefaultendpunct}{\mcitedefaultseppunct}\relax
\EndOfBibitem
\bibitem{BaBar_pph1}
BaBar collaboration, B.~Aubert {\em et~al.},
  \ifthenelse{\boolean{articletitles}}{{\it {Measurement of the $B^+\to
  p\bar{p}K^+$ branching fraction and study of the decay dynamics}},
  }{}\href{http://dx.doi.org/10.1103/PhysRevD.72.051101}{Phys.\ Rev.\  {\bf
  D72} (2005) 051101}, \href{http://arxiv.org/abs/hep-ex/0507012}{{\tt
  arXiv:hep-ex/0507012}}\relax
\mciteBstWouldAddEndPuncttrue
\mciteSetBstMidEndSepPunct{\mcitedefaultmidpunct}
{\mcitedefaultendpunct}{\mcitedefaultseppunct}\relax
\EndOfBibitem
\bibitem{BaBar_pph2}
BaBar collaboration, B.~Aubert {\em et~al.},
  \ifthenelse{\boolean{articletitles}}{{\it {Evidence for the $B^0\to
  p\bar{p}K^{*0}$ and $B^+\to\etac K^{*+}$ decays and study of the decay
  dynamics of B meson decays into $p\bar{p}h$ final states}},
  }{}\href{http://dx.doi.org/10.1103/PhysRevD.76.092004}{Phys.\ Rev.\  {\bf
  D76} (2007) 092004}, \href{http://arxiv.org/abs/0707.1648}{{\tt
  arXiv:0707.1648}}\relax
\mciteBstWouldAddEndPuncttrue
\mciteSetBstMidEndSepPunct{\mcitedefaultmidpunct}
{\mcitedefaultendpunct}{\mcitedefaultseppunct}\relax
\EndOfBibitem
\bibitem{Belle_pph}
Belle collaboration, J.-T. Wei {\em et~al.},
  \ifthenelse{\boolean{articletitles}}{{\it {Study of the decay mechanism for
  $B^+\to p\bar{p}K^+$ and $B^+\to p\bar{p}\pi^+$}},
  }{}\href{http://dx.doi.org/10.1016/j.physletb.2007.11.063}{Phys.\ Lett.\ B
  {\bf 659} (2008) 80}, \href{http://arxiv.org/abs/0706.4167}{{\tt
  arXiv:0706.4167}}\relax
\mciteBstWouldAddEndPuncttrue
\mciteSetBstMidEndSepPunct{\mcitedefaultmidpunct}
{\mcitedefaultendpunct}{\mcitedefaultseppunct}\relax
\EndOfBibitem
\bibitem{LHCb-PAPER-2012-047}
LHCb collaboration, R.~Aaij {\em et~al.},
  \ifthenelse{\boolean{articletitles}}{{\it {Measurements of the branching
  fractions of $B^+ \to p\bar{p}K^+$ decays}},
  }{}\href{http://dx.doi.org/10.1140/epjc/s10052-013-2462-2}{Eur.\ Phys.\ J.\
  {\bf C73} (2013) 2462}, \href{http://arxiv.org/abs/1303.7133}{{\tt
  arXiv:1303.7133}}\relax
\mciteBstWouldAddEndPuncttrue
\mciteSetBstMidEndSepPunct{\mcitedefaultmidpunct}
{\mcitedefaultendpunct}{\mcitedefaultseppunct}\relax
\EndOfBibitem
\bibitem{Alves:2008td}
LHCb collaboration, A.~A. Alves~Jr. {\em et~al.},
  \ifthenelse{\boolean{articletitles}}{{\it {The LHCb detector at the LHC}},
  }{}\href{http://dx.doi.org/10.1088/1748-0221/3/08/S08005/}{JINST {\bf 3}
  (2008) S08005}\relax
\mciteBstWouldAddEndPuncttrue
\mciteSetBstMidEndSepPunct{\mcitedefaultmidpunct}
{\mcitedefaultendpunct}{\mcitedefaultseppunct}\relax
\EndOfBibitem
\bibitem{RichPerf}
M.~Adinolfi {\em et~al.}, \ifthenelse{\boolean{articletitles}}{{\it
  {Performance of the \lhcb RICH detector at the LHC}}, }{}Eur.\ Phys.\ J.\
  {\bf C73} (2013) 2431, \href{http://arxiv.org/abs/1211.6759}{{\tt
  arXiv:1211.6759}}\relax
\mciteBstWouldAddEndPuncttrue
\mciteSetBstMidEndSepPunct{\mcitedefaultmidpunct}
{\mcitedefaultendpunct}{\mcitedefaultseppunct}\relax
\EndOfBibitem
\bibitem{Aaij:2012me}
R.~Aaij {\em et~al.}, \ifthenelse{\boolean{articletitles}}{{\it {The LHCb
  trigger and its performance}},
  }{}\href{http://dx.doi.org/10.1088/1748-0221/8/04/P04022}{JINST {\bf 8}
  (2013) P04022}, \href{http://arxiv.org/abs/1211.3055}{{\tt
  arXiv:1211.3055}}\relax
\mciteBstWouldAddEndPuncttrue
\mciteSetBstMidEndSepPunct{\mcitedefaultmidpunct}
{\mcitedefaultendpunct}{\mcitedefaultseppunct}\relax
\EndOfBibitem
\bibitem{Gligorov:2012qt}
V.~V. Gligorov and M.~Williams, \ifthenelse{\boolean{articletitles}}{{\it
  {Efficient, reliable and fast high-level triggering using a bonsai boosted
  decision tree}},
  }{}\href{http://dx.doi.org/10.1088/1748-0221/8/02/P02013}{JINST {\bf 8}
  (2013) P02013}, \href{http://arxiv.org/abs/1210.6861}{{\tt
  arXiv:1210.6861}}\relax
\mciteBstWouldAddEndPuncttrue
\mciteSetBstMidEndSepPunct{\mcitedefaultmidpunct}
{\mcitedefaultendpunct}{\mcitedefaultseppunct}\relax
\EndOfBibitem
\bibitem{Sjostrand:2006za}
T.~Sj\"{o}strand, S.~Mrenna, and P.~Skands,
  \ifthenelse{\boolean{articletitles}}{{\it {PYTHIA 6.4 physics and manual}},
  }{}\href{http://dx.doi.org/10.1088/1126-6708/2006/05/026}{JHEP {\bf 05}
  (2006) 026}, \href{http://arxiv.org/abs/hep-ph/0603175}{{\tt
  arXiv:hep-ph/0603175}}\relax
\mciteBstWouldAddEndPuncttrue
\mciteSetBstMidEndSepPunct{\mcitedefaultmidpunct}
{\mcitedefaultendpunct}{\mcitedefaultseppunct}\relax
\EndOfBibitem
\bibitem{LHCb-PROC-2010-056}
I.~Belyaev {\em et~al.}, \ifthenelse{\boolean{articletitles}}{{\it {Handling of
  the generation of primary events in \gauss, the \lhcb simulation framework}},
  }{}\href{http://dx.doi.org/10.1109/NSSMIC.2010.5873949}{Nuclear Science
  Symposium Conference Record (NSS/MIC) {\bf IEEE} (2010) 1155}\relax
\mciteBstWouldAddEndPuncttrue
\mciteSetBstMidEndSepPunct{\mcitedefaultmidpunct}
{\mcitedefaultendpunct}{\mcitedefaultseppunct}\relax
\EndOfBibitem
\bibitem{Lange:2001uf}
D.~J. Lange, \ifthenelse{\boolean{articletitles}}{{\it {The EvtGen particle
  decay simulation package}},
  }{}\href{http://dx.doi.org/10.1016/S0168-9002(01)00089-4}{Nucl.\ Instrum.\
  Meth.\  {\bf A462} (2001) 152}\relax
\mciteBstWouldAddEndPuncttrue
\mciteSetBstMidEndSepPunct{\mcitedefaultmidpunct}
{\mcitedefaultendpunct}{\mcitedefaultseppunct}\relax
\EndOfBibitem
\bibitem{Golonka:2005pn}
P.~Golonka and Z.~Was, \ifthenelse{\boolean{articletitles}}{{\it {PHOTOS Monte
  Carlo: a precision tool for QED corrections in $Z$ and $W$ decays}},
  }{}\href{http://dx.doi.org/10.1140/epjc/s2005-02396-4}{Eur.\ Phys.\ J.\  {\bf
  C45} (2006) 97}, \href{http://arxiv.org/abs/hep-ph/0506026}{{\tt
  arXiv:hep-ph/0506026}}\relax
\mciteBstWouldAddEndPuncttrue
\mciteSetBstMidEndSepPunct{\mcitedefaultmidpunct}
{\mcitedefaultendpunct}{\mcitedefaultseppunct}\relax
\EndOfBibitem
\bibitem{Allison:2006ve}
Geant4 collaboration, J.~Allison {\em et~al.},
  \ifthenelse{\boolean{articletitles}}{{\it {Geant4 developments and
  applications}}, }{}\href{http://dx.doi.org/10.1109/TNS.2006.869826}{IEEE
  Trans.\ Nucl.\ Sci.\  {\bf 53} (2006) 270}\relax
\mciteBstWouldAddEndPuncttrue
\mciteSetBstMidEndSepPunct{\mcitedefaultmidpunct}
{\mcitedefaultendpunct}{\mcitedefaultseppunct}\relax
\EndOfBibitem
\bibitem{Agostinelli:2002hh}
Geant4 collaboration, S.~Agostinelli {\em et~al.},
  \ifthenelse{\boolean{articletitles}}{{\it {Geant4: a simulation toolkit}},
  }{}\href{http://dx.doi.org/10.1016/S0168-9002(03)01368-8}{Nucl.\ Instrum.\
  Meth.\  {\bf A506} (2003) 250}\relax
\mciteBstWouldAddEndPuncttrue
\mciteSetBstMidEndSepPunct{\mcitedefaultmidpunct}
{\mcitedefaultendpunct}{\mcitedefaultseppunct}\relax
\EndOfBibitem
\bibitem{LHCb-PROC-2011-006}
M.~Clemencic {\em et~al.}, \ifthenelse{\boolean{articletitles}}{{\it {The \lhcb
  simulation application, \gauss: design, evolution and experience}},
  }{}\href{http://dx.doi.org/10.1088/1742-6596/331/3/032023}{{J.\ Phys.\ \!\!:
  Conf.\ Ser.\ } {\bf 331} (2011) 032023}\relax
\mciteBstWouldAddEndPuncttrue
\mciteSetBstMidEndSepPunct{\mcitedefaultmidpunct}
{\mcitedefaultendpunct}{\mcitedefaultseppunct}\relax
\EndOfBibitem
\bibitem{CLEO_pp}
CLEO collaboration, S.~B. Athar {\em et~al.},
  \ifthenelse{\boolean{articletitles}}{{\it {Radiative decays of the
  $\Upsilon(1S)$ to a pair of charged hadrons}},
  }{}\href{http://dx.doi.org/10.1103/PhysRevD.73.032001}{Phys.\ Rev.\  {\bf
  D73} (2006) 032001}, \href{http://arxiv.org/abs/hep-ex/0510015}{{\tt
  arXiv:hep-ex/0510015}}\relax
\mciteBstWouldAddEndPuncttrue
\mciteSetBstMidEndSepPunct{\mcitedefaultmidpunct}
{\mcitedefaultendpunct}{\mcitedefaultseppunct}\relax
\EndOfBibitem
\bibitem{BES_pp}
BES collaboration, J.~Z. Bai {\em et~al.},
  \ifthenelse{\boolean{articletitles}}{{\it {Observation of a near-threshold
  enhancement in the $p\bar{p}$ mass spectrum from radiative $\jpsi\to\gamma
  p\bar{p}$}}, }{}\href{http://dx.doi.org/10.1103/PhysRevLett.91.022001}{Phys.\
  Rev.\ Lett {\bf 91} (2003) 022001},
  \href{http://arxiv.org/abs/hep-ex/0303006}{{\tt arXiv:hep-ex/0303006}}\relax
\mciteBstWouldAddEndPuncttrue
\mciteSetBstMidEndSepPunct{\mcitedefaultmidpunct}
{\mcitedefaultendpunct}{\mcitedefaultseppunct}\relax
\EndOfBibitem
\bibitem{BaBar_dpp}
BaBar collaboration, B.~Aubert {\em et~al.},
  \ifthenelse{\boolean{articletitles}}{{\it {Measurements of the decays
  $B^0\to\bar{D}^0p\bar{p}$, $B^0\to\bar{D}^{*0}p\bar{p}$, $B^0\to
  D^-p\bar{p}\pi^+$ and $B^0\to D^{*-}p\bar{p}\pi^+$}},
  }{}\href{http://dx.doi.org/10.1103/PhysRevD.74.051101}{Phys.\ Rev.\  {\bf
  D74} (2006) 051101}, \href{http://arxiv.org/abs/hep-ex/0607039}{{\tt
  arXiv:hep-ex/0607039}}\relax
\mciteBstWouldAddEndPuncttrue
\mciteSetBstMidEndSepPunct{\mcitedefaultmidpunct}
{\mcitedefaultendpunct}{\mcitedefaultseppunct}\relax
\EndOfBibitem
\bibitem{Haidenbauer}
J.~Haidenbauer {\em et~al.}, \ifthenelse{\boolean{articletitles}}{{\it {Near
  threshold $p\bar{p}$ enhancement in $B$ and \jpsi decays}},
  }{}\href{http://dx.doi.org/10.1103/PhysRevD.74.017501}{Phys.\ Rev.\  {\bf
  D74} (2006) 017501}, \href{http://arxiv.org/abs/hep-ph/0605127}{{\tt
  arXiv:hep-ph/0605127}}\relax
\mciteBstWouldAddEndPuncttrue
\mciteSetBstMidEndSepPunct{\mcitedefaultmidpunct}
{\mcitedefaultendpunct}{\mcitedefaultseppunct}\relax
\EndOfBibitem
\bibitem{HYCheng}
H.~Y. Cheng, \ifthenelse{\boolean{articletitles}}{{\it {Exclusive baryonic $B$
  decays circa 2005}},
  }{}\href{http://dx.doi.org/10.1142/S0217751X06033969}{Int.\ J.\ Mod.\ Phys.\
  {\bf A21} (2006) 4209}, \href{http://arxiv.org/abs/hep-ph/0603003}{{\tt
  arXiv:hep-ph/0603003}}\relax
\mciteBstWouldAddEndPuncttrue
\mciteSetBstMidEndSepPunct{\mcitedefaultmidpunct}
{\mcitedefaultendpunct}{\mcitedefaultseppunct}\relax
\EndOfBibitem
\bibitem{sPlots_NIM}
M.~{Pivk} and F.~R. {Le Diberder}, \ifthenelse{\boolean{articletitles}}{{\it
  {sPlot: A statistical tool to unfold data distributions}},
  }{}\href{http://dx.doi.org/10.1016/j.nima.2005.08.106}{Nucl.\ Instrum.\
  Meth.\  {\bf A555} (2005) 356},
  \href{http://arxiv.org/abs/physics/0402083}{{\tt
  arXiv:physics/0402083}}\relax
\mciteBstWouldAddEndPuncttrue
\mciteSetBstMidEndSepPunct{\mcitedefaultmidpunct}
{\mcitedefaultendpunct}{\mcitedefaultseppunct}\relax
\EndOfBibitem
\bibitem{PDG}
Particle Data Group, J.~Beringer {\em et~al.},
  \ifthenelse{\boolean{articletitles}}{{\it {\href{http://pdg.lbl.gov/}{Review
  of particle physics}}},
  }{}\href{http://dx.doi.org/10.1103/PhysRevD.86.010001}{Phys.\ Rev.\  {\bf
  D86} (2012) 010001}\relax
\mciteBstWouldAddEndPuncttrue
\mciteSetBstMidEndSepPunct{\mcitedefaultmidpunct}
{\mcitedefaultendpunct}{\mcitedefaultseppunct}\relax
\EndOfBibitem
\bibitem{Saphir}
F.~W. Wieland {\em et~al.}, \ifthenelse{\boolean{articletitles}}{{\it {Study of
  the reaction $\gamma p \to K^+\Lz(1520)$ at photon energies up to
  2.65~\gev}}, }{}\href{http://dx.doi.org/10.1140/epja/i2011-11047-x}{Eur.\
  Phys.\ J.\  {\bf A47} (2011) 47, erratum ibid: \textbf{A47} (2011) 133},
  \href{http://arxiv.org/abs/1011.0822}{{\tt arXiv:1011.0822}}\relax
\mciteBstWouldAddEndPuncttrue
\mciteSetBstMidEndSepPunct{\mcitedefaultmidpunct}
{\mcitedefaultendpunct}{\mcitedefaultseppunct}\relax
\EndOfBibitem
\bibitem{Laporta}
V.~Laporta, \ifthenelse{\boolean{articletitles}}{{\it {Final state interaction
  enhancement effect on the near threshold $p\bar{p}$ system in the $B^{\pm}\to
  p\bar{p}\pi^{\pm}$ decay}},
  }{}\href{http://dx.doi.org/10.1142/S0217751X07037949}{Int.\ J.\ Mod.\ Phys.\
  {\bf A22} (2007) 5401}, \href{http://arxiv.org/abs/0707.2751}{{\tt
  arXiv:0707.2751}}\relax
\mciteBstWouldAddEndPuncttrue
\mciteSetBstMidEndSepPunct{\mcitedefaultmidpunct}
{\mcitedefaultendpunct}{\mcitedefaultseppunct}\relax
\EndOfBibitem
\end{mcitethebibliography}
\end{document}